\documentclass[twocolumn,superscriptaddress,amsmath,amssymb,aps]{revtex4-2}
\usepackage{hyperref}
\usepackage{color}
\usepackage{graphicx}
\usepackage{subfigure}
\usepackage{booktabs}
\usepackage{dcolumn}
\usepackage{braket}
\usepackage{bm}
\usepackage{amsmath,mathtools}
\usepackage{verbatim}
\usepackage{lineno}
\usepackage[T1]{fontenc}
\usepackage{threeparttable}

\newcommand{\beq}{\begin{eqnarray}}
\newcommand{\eeq}{\end{eqnarray}}

\newcommand{\red}[1]{\textcolor{red}{#1}}

\newcommand\YJ[1]{\textcolor{red}{ [YJ:\,#1]}}

\begin{document}



\title{Universal activated aging and weak ergodicity breaking in spin and structural glasses}

\author{Bin Li}
\affiliation{Institute of Theoretical Physics, Chinese Academy of Sciences, Beijing 100190, China}
\affiliation{School of Physical Sciences, University of Chinese Academy of Sciences, Beijing 100049, China}

\author{Deng Pan}
\affiliation{Institute of Theoretical Physics, Chinese Academy of Sciences, Beijing 100190, China}

\author{Ting Qu}
\affiliation{Institute of Theoretical Physics, Chinese Academy of Sciences, Beijing 100190, China}
\affiliation{School of Physical Sciences, University of Chinese Academy of Sciences, Beijing 100049, China}


\author{Yuliang Jin}
\email{yuliangjin@mail.itp.ac.cn}
\affiliation{Institute of Theoretical Physics, Chinese Academy of Sciences, Beijing 100190, China}
\affiliation{School of Physical Sciences, University of Chinese Academy of Sciences, Beijing 100049, China}
\affiliation{Center for Theoretical Interdisciplinary Sciences, Wenzhou Institute, University of Chinese Academy of Sciences, Wenzhou, Zhejiang 325001, China}

\date{\today}
\begin{abstract}
Glasses possess complex energy landscapes and exhibit non-equilibrium aging dynamics.  Here, we propose a generalized trap model for activated aging  based on a key static property of the energy landscape: the distribution of energy barriers. Our theory predicts that, upon cooling, weak ergodicity breaking (WEB) in  quenching dynamics occurs prior to strong ergodicity breaking in equilibrium dynamics.
Furthermore, the theory indicates that the characteristic size of activation clusters can be deduced from the logarithmic decay of the time-correlation function.
We rigorously test the model's assumptions and predictions using the  simplest spin glass model - the random energy model. The predicted aging behavior is also universally observed in paradigmatic structural glasses, including the Weeks-Chandler-Andersen (WCA) model and amorphous silica. Remarkably, applying our framework to the WCA model allows us to extract a static  length from the non-equilibrium dynamics, extending its observable growth range from a mere factor of 2-3 to a full order of magnitude and providing supportive evidence for the random first-order transition scenario.  Finally, we propose a unified ergodic-WEB phase diagram for aging 
 dynamics in general glassy systems. 
\end{abstract}

\maketitle

\section{Introduction}
Glassy systems, including spin glasses~\cite{mezard1987spin}, structural glasses~\cite{parisi2020theory}, polymers~\cite{struik1978physical}, colloidal suspensions~\cite{courtland2002direct}, granular materials~\cite{gago2020universal}, active matter~\cite{janzen2022aging} and artificial neural networks~\cite{baity2018comparing}, are characterized by multiple thermodynamically metastable states (glass basins) that are separated by energy barriers in the energy landscape~\cite{stillinger2015energy}, and non-equilibrium aging dynamics~\cite{bouchaud1997aging, Cugliandolo2003Dynamics, arceri2022glasses}. 
Aging refers to the dynamical slowing down of relaxation processes with an increasing ``age''  quantified by the waiting time (or the aging time) $t_{\rm w}$ elapsed after the system is quenched. Understanding aging represents a major theoretical challenge. Three approaches have been developed. 

\begin{figure*}[!htbp]
  \centering 
  \includegraphics[width=\linewidth]{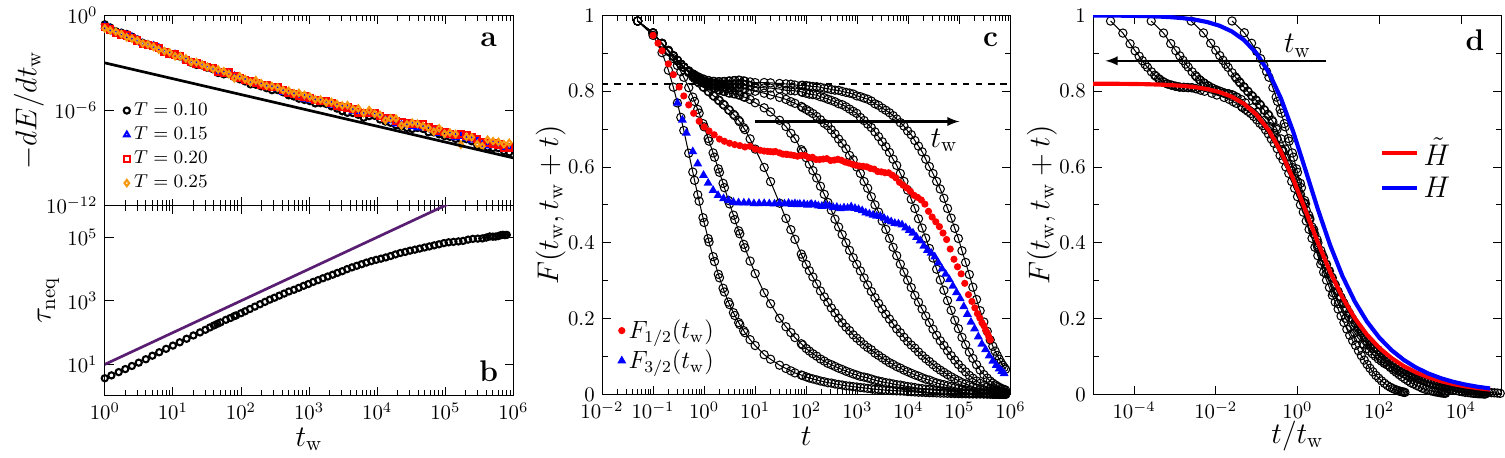}
  \caption{{\bf  Basic aging phenomenon in glasses.
  } Data are obtained for the WCA model at $T=0.25$ with  $N=8000$ particles, quenched from $T_{\rm eq} = 5$ to $T$. 
  (a) $-dE/dt_{\rm w}$ as a function of $t_{\rm w}$ at a few different $T$. The solid line indicates $-dE/dt_{\rm w} \sim t_{\rm w}^{-1}$, i.e., $E(t_{\rm w}) \sim - \ln(t_{\rm w})$.
  (b) Non-equilibrium decorrelation time $\tau_{\rm neq}$ as a function of $t_{\rm w}$. The solid line represents $\tau_{\rm neq} \sim t_{\rm w}$.
  (c) $F(t_{\rm w}, t_{\rm w}+t)$ for $t_{\rm w} = 0.2, 2, 20, 2\times 10^2, 2\times 10^3, 2\times 10^4, 8\times 10^5$ (black curves; from left to right),  and $F_w(t_{\rm w})$ with $w=1/2$ and $w=3/2$. The dashed line indicates the nonergodicity parameter $f$.
  (c) $F(t_{\rm w}, t_{\rm w}+t)$ plotted as a function of $w = t/t_{\rm w}$ for intermediate waiting times $t_{\rm w} = 2, 20, 2\times 10^2, 2\times 10^3$.
  The blue line is $H$ given by Eq.~(\ref{eq:Pi}) at the given $\hat{x}=T/T_{\rm WEB}$, and the red line is $\tilde{H} =  f H$.
  }
  \label{fig:aging}
\end{figure*}

{\it (i) Mean-field theories.} 
In mean-field glass models, aging is described by a set of closed equations for the two-time correlation and response functions~\cite{cugliandolo1993analytical, cugliandolo1996large, altieri2020dynamical, cugliandolo1994out, bernaschi2020strong, folena2020rethinking, folena2023weak}. Mean-field aging corresponds to slow descent in the  (free) energy landscape after quenching, during which paths to find lower energies become more and more scarce -- called an ``entropic effect''~\cite{kurchan1996phase, bouchaud1997aging, bouchaud1998out, agoritsas2019out}. Because energy barriers are infinite in mean-field models (in the thermodynamic limit), activated barrier-crossing processes do not occur in mean-field aging. Efforts to go beyond the mean-field approximation proceed mainly in the following two directions.

{\it (ii) Phenomenological trap models: random walk in  a complex energy landscape.} 
Trap models~\cite{bouchaud1992weak, bouchaud1995aging} are based on an intuitive picture that the evolution of a glass system can be described by the dynamics  of a random walker in a complex multidimensional  energy landscape~\cite{goldstein1969viscous, debenedetti2001supercooled}. The random walker is typically ``trapped'' in a metastable glass basin (trap), within a trapping time  $\tau$, whose distribution is determined by the property of the landscape. With an increasing $t_{\rm w}$, the probability of the system being trapped in deeper basins increases, which results in a longer hopping time to escape the basin and thus slower relaxation dynamics.  

{\it (iii) Coarsening mechanism in real space.} The aging effects have also been analyzed during coarsening processes  in the paradigmatic Ising model~\cite{bray1994theory}, models with quenched interaction disorder or random field~\cite{fisher1986ordered, bray1984lower}, and kinetically constrained models with facilitated dynamics~\cite{mayer2007ageing}. 
Aging is generally characterized by logarithmic growth of the domain size.

To understand aging in real glassy systems, it is inevitable to go beyond mean-field approximations. The landscape-based approach, which we follow in the current study,  is particularly attractive  due to its simplicity and deep connections to general stochastic processes modeled by  continuous-time random walk ~\cite{bouchaud1990anomalous, li2025route}. 
The trap model is initially proposed by  Bouchaud (BTM)~\cite{bouchaud1992weak, bouchaud1995aging} to describe aging dynamics in the simplest spin glass model~\cite{gross1984simplest}, i.e., the random energy model (REM)~\cite{derrida1980random}. 
An inspiring  outcome from the BTM is the so-called weak ergodicity breaking (WEB) in the spin glass phase, where the non-equilibrium correlation function converges, in the large-time limit, to a  plateau  whose value is determined by the so-called arcsin law~\cite{bouchaud1995aging}.  
Such  asymptotic solutions of the BTM are supported by recent rigorous mathematical  derivations~\cite{vcerny2017aging, gayrard2019aging, gayrard2019dynamic, derrida2023random}. 
 However, this  result is inconsistent with  Monte Carlo (MC) simulations of the REM (with Gaussian random energies) consisting of a finite number $N$ of spins ~\cite{baity2018activated, stariolo2019activated}: the obtained aging functions do not show any sign of convergence to the predicted plateau. Obviously, before any rigorous extension can be made, a fundamental task is to reconcile the inconsistency between the theory and MC simulations in the REM.  Once this mission is accomplished, the next natural question is about the relevance of trap model predictions to the aging behavior in more realistic glasses~\cite{denny_trap_2002}. Both problems are addressed in the present study.

Here we propose a generalized trap model (GTM), which takes into account the finite-size corrections in the barrier energy distribution. The GTM theory predicts
that the WEB in non-equilibrium aging  occurs at a temperature $T_{\rm WEB}$, above the strong (standard) ergodicity breaking (SEB) temperature $T_{\rm SEB}$ in equilibrium dynamics. This WEB corresponds to the point where the large-waiting-time limit and large-system-size limit (thermodynamic limit) become non-interchangeable.  The theory also explains that the logarithmic decay in the time-correlation aging function is associated with the finite spatial size of activated events. The GTM predictions are 
consistent with simulation results of four models: random energy spin glass models with Gaussian and exponential energy distributions, Weeks-Chandler-Andersen
(WCA) structural glass model, and amorphous silica ($\rm{SiO}_2$) network glass model. 
Remarkably, combining the theoretical formalism and simulation data, a static length $\xi_{\rm ag}$ characterizing the size of activation patches can be extracted from non-equilibrium aging dynamics in the WCA model. The $\xi_{\rm ag}$ data, although lying in a much lower temperature regime, are on the same curve of other static lengths, including the point-to-set (PTS) length  $\xi_{\rm PTS}$~\cite{bouchaud2004adam, hocky2012growing} and the static length $\xi_{\rm Hessian}$ obtained from vibrational properties~\cite{karmakar2012direct}. Combining the three lengths together, the growth range of the static length is extended from a factor of $2\sim3$ existing in the literature to one decade, showing supporting evidence of the  random first order transition (RFOT) prediction $\xi \sim (T-T_{\rm K})^{-\nu}$ with a finite Kauzmann temperature $T_{\rm K}$ and  $\nu \approx 0.61$~\cite{kirkpatrick1989scaling, lubchenko2007theory}. Our results are summarized by a phase diagram that can universally describe ergodic and WEB phases in spin and structural glasses. \\

\section{Basic aging phenomenon and weak ergodicity breaking}
The aging phenomenon in glasses is typically studied as follows. Starting from an equilibrium configuration at a high temperature $T_{\rm eq}$, one rapidly quenches the system to a final temperature $T < T_{\rm eq}$, and lets the system evolve at a constant temperature $T$ (the time is set to zero $t=0$ after quenching).  The system is out of equilibrium after quenching, and the following relaxation dynamics are accompanied by aging effects. 

The aging dynamics can be monitored by one-time observables, such as the potential energy $E(t_{\rm w})$, which typically shows slow decay close to a logarithmic time-dependent function, $E(t_{\rm w}) \sim - \ln(t_{\rm w})$ (see Fig.~\ref{fig:aging}a; to be explicit, in Fig.~\ref{fig:aging} we focus on  the  simulation results of the WCA model). 
More informative descriptors are two-time correlation functions, $C(t_{\rm w}, t_{\rm w} + t) = \langle A^*(t_{\rm w}) A(t_{\rm w}+t) \rangle$,
where $A$ is a physical observable, $t_{\rm w}$ the {waiting time} elapsed after quenching, and $\langle \ldots \rangle$ the average over initial configurations. In structural glasses, one typically takes $A(t) = e^{i {\bf q} \cdot {\bf r}_i(t)}$, where ${\bf r}_i(t)$ is the position of particle $i$ at time $t$, and ${\bf q}$ is a wave-vector - then $F_q (t_{\rm w}, t_{\rm w} + t) = \frac{1}{N} \langle \sum_{i=1}^N e^{i {\bf q} \cdot [{\bf r}_i(t_{\rm w}+t) -  {\bf r}_i(t_{\rm w})]} \rangle$ is essentially the non-equilibrium generalization of the standard incoherent scattering function. In spin glasses, it is natural to consider the spin-spin correlations, $C(t_{\rm w}, t_{\rm w}+t) = \frac{1}{N}\langle \sum_{i=1}^N S_i(t_{\rm w}) S_i(t_{\rm w}+t) \rangle $. Here $N$ represents the system size, i.e., number of particles or spins. Figure~\ref{fig:aging}b shows that  $F_q (t_{\rm w}, t_{\rm w} + t)$ in the WCA model has a strong dependence on $t_{\rm w}$, which is direct evidence of non-equilibrium dynamics. Following the convention, in this study we focus on a fixed $q=q_{\rm max}$, where $q_{\rm max}$ is the location of the maximum in the structural factor $S(q)$. The subscript of $q$   in $F_q$ will be omitted from now on.

\begin{figure*}[!htbp]
  \centering
  \includegraphics[width=0.7\linewidth]{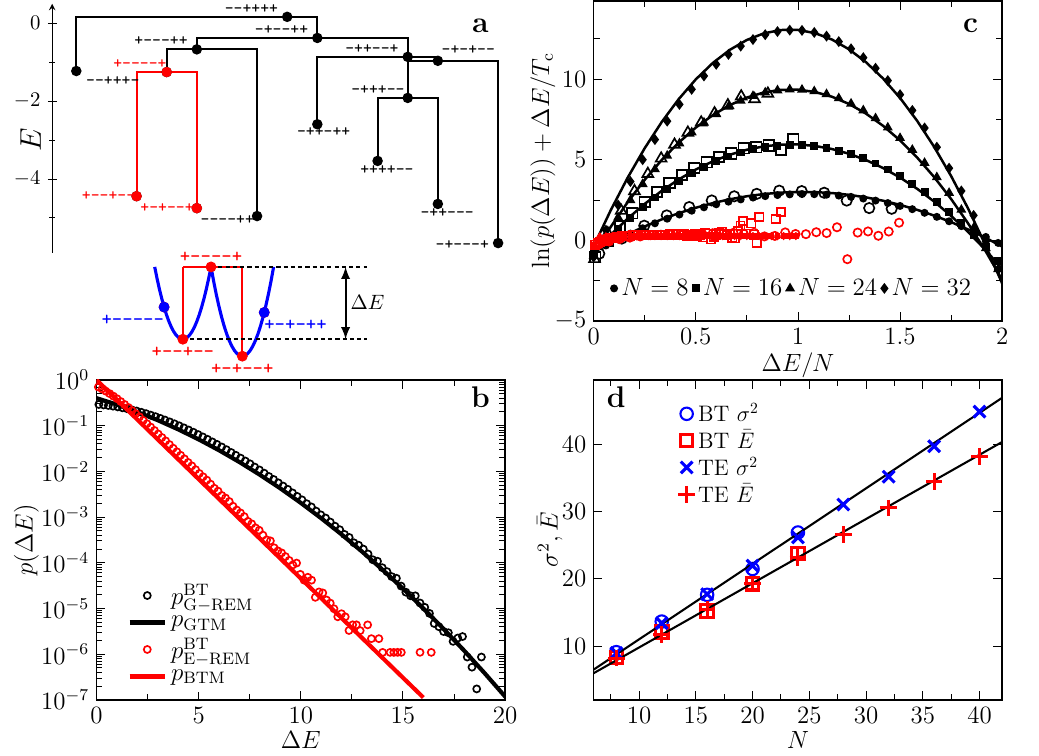}
  \caption{{\bf Barrier energy distributions in the G-REM and E-REM.}
  (a) A sub-tree formed by local minima and saddle points for the $N=6$ G-REM obtained by the BT algorithm, with corresponding spin configurations and energies ($E$-axis) indicated. (inset) Illustration of two traps: the blue nodes are non-minimum-non-saddle configurations;
  the barrier energy $\Delta E$ is the energy difference between the local minimum and saddle point. 
(b) Distributions $p_{\rm G-REM}^{\rm BT}(\Delta E)$ 
and $p_{\rm E-REM}^{\rm BT}(\Delta E)$ of the G-REM and E-REM 
by the BT algorithm ($N=20$),
compared to $p_{\rm BTM}(\Delta E)$ (Eq.~\ref{eq:BTM}) and $p_{\rm GTM}(\Delta E)$ (Eq.~\ref{eq:GTM} with $a=1.12$ and $b=0.96$).
(c) Remainder $\ln p(\Delta E) + \Delta E/T_{\rm c}$ for the  G-REM (black) and E-REM (red). 
 Open and filled points are obtained by the BT algorithm and TE theory respectively.
Lines represent fitting to a Gaussian function. The variance $\sigma^2$ and mean  $\bar{E}$ obtained by the Gaussian fitting are plotted in (d) as functions of $N$. Linear fitting of the data in (d) gives $a=1.12$ and $b=0.96$ ($\sigma^2=aN$ and $\bar{E}=bN$).}
  \label{fig:barrier}
\end{figure*}

It can be argued that, in many cases, the key object describing  aging is the reduced one-time correlation function ({\it aging function}), $C_w(t_{\rm w}) \equiv C[t_{\rm w}, (1+w) t_{\rm w}] $
with the ratio $t/t_{\rm w} = w$  fixed. This is based on the following considerations. First, aging effects are only observable after a time scale in the order of $t_{\rm w}$ - at short times, the correlation functions $F (t_{\rm w}, t_{\rm w} + t)$ is $t_{\rm w}$-independent (see Fig.~\ref{fig:aging}c)~\cite{kob1997aging}. Second, if plotted as a function of $t/t_{\rm w}$, $F (t_{\rm w}, t_{\rm w} + t)$ satisfies  $F (t_{\rm w}, t_{\rm w} + t) = F^{\rm st}(t) + F^{\rm ag}(t/t_{\rm w})$  for intermediate $t_{\rm w}$ (see Fig.~\ref{fig:aging}d). 
Here the first term $F^{\rm st}(t)$ corresponds to short-time microscopic dynamics that are $t_{\rm w}$-independent. The collapsed term $F^{\rm ag}(t/t_{\rm w})$ is the aging part that we aim to understand.
Note that curves with very small $t_{\rm w}$ and very large $t_{\rm w}$, corresponding to microscopic and equilibrium dynamics respectively, do not collapse.
The non-equilibrium decorrelation time, defined by $F (t_{\rm w}, t_{\rm w} + \tau_{\rm neq}) =  1/e$, is nearly proportional to $t_{\rm w}$ in the aging regime, $\tau_{\rm neq}(t_{\rm w})\sim t_{\rm w}$ (see Fig.~\ref{fig:aging}b).
At large times, $\tau_{\rm neq}(t_{\rm w})$ reaches a plateau, $\tau_{\rm neq}(t_{\rm w} \to \infty) = \tau_{\rm eq}$, due to thermalization. 
Although a more general rescaling form has been proposed based on the coarsening mechanism, 
$C(t_{\rm w}, t_{\rm w} + t) = C^{\rm st}(t) + C^{\rm ag}[h(t+t_{\rm w})/h(t_{\rm w})]$~\cite{bray1994theory}, here we will only consider the simplistic case, $h(t) \sim t$, which already captures well the simulation results. Third, $C_w(t_{\rm w})$ can be theoretically analyzed by trap models, with asymptotic behavior given by the arcsin law~\cite{bouchaud1992weak, bouchaud1995aging, bouchaud1997aging}.

A remarkable property associated  with $C_w(t_{\rm w})$ is the WEB behavior, originally noticed by Bouchaud in the analysis of the BTM ~\cite{ bouchaud1992weak, cugliandolo1993analytical, bouchaud1995aging, folena2023weak}. One is interested in the following question: if the system is quenched from $T_{\rm eq}$ to a low temperature $T$ and waited for a time $t_{\rm w}$, can it relax after a time $t = w \, t_{\rm w}$? Mathematically, the WEB is defined by the condition when $C_w(t_{\rm w})$ does not decay to zero in the large-$t_{\rm w}$ limit: $C_w(t_{\rm w} \to \infty) > 0$ for $T<T_{\rm WEB}$, and $C_w(t_{\rm w} \to \infty) = 0$ for $T> T_{\rm WEB}$, where $T_{\rm WEB}$ is the WEB temperature.  Such WEB is observable in the simulated $F_w(t_{\rm w}) \equiv F[t_{\rm w}, (1+w) t_{\rm w}]$ with a fixed $w$ at $T=0.25$ (see Fig.~\ref{fig:aging}c). The plateau in the intermediate time ($10^0<t_{\rm w}<10^4$) is evidence of WEB.
The decay from this plateau at large $t_{\rm w}$ is due to thermalization, which is not considered by the current theoretical model.

Intuitively, the WEB indicates the detection of the energy landscape in non-equilibrium aging dynamics when $T<T_{\rm WEB}$. The system ``feels'' the energy landscape only when the
time scales of fast and activated dynamics are separable. 
In this case,  according to the Arrhenius law, the system is typically trapped in a metastable glass basin with an energy barrier $\Delta E \sim k_{\rm B} T \ln (t_{\rm w})$ after a waiting time $t_{\rm w}$ (we will neglect the Boltzmann constant $k_{\rm B}=1$ in the expressions below by setting it as the unit). The time required to escape this basin is $\tau \sim e^{\Delta E/T} \sim t_{\rm w}$. In other words, whenever the system feels the landscape, ergodicity breaks down within the timescale $t \sim t_{\rm w}$. However, such ergodicity breaking is ``weak'' in the sense that the system relaxes after a time scale $t$ much larger than $t_{\rm w}$. Indeed,  for a finite, fixed $t_{\rm w}$, $\lim_{t \to \infty} F(t_{\rm w}, t_{\rm w} + t) = 0$ at the same $T$. As shown by Fig.~\ref{fig:aging}c, in the time regime $10^0<t_{\rm w}<10^4$  where $F_w(t_{\rm w})$ exhibits a plateau, the two-time function $F(t_{\rm w}, t_{\rm w} + t)$ for a fixed  $t_{\rm w}$ rapidly decays with an increasing $t$,  showing no sign of a plateau.  

The WEB in non-equilibrium aging dynamics is, by definition, different from the SEB. 
The SEB occurs at a temperature $T_{\rm SEB}$ where the phase space becomes disconnected. 
Mathematically, the SEB is described  by  the equilibrium time-correlation function, $C_{\rm eq}(t) = \langle  C_{\rm eq}(t_0, t_0+t) \rangle_{t_0}$, where $\langle  \ldots \rangle_{t_0}$ is the average over the reference time $t_0$ based on the time translation invariance: $C_{\rm eq}(t \to \infty) > 0$ for $T<T_{\rm SEB}$, and  $C_{\rm eq}(t \to \infty) = 0$ for $T > T_{\rm SEB}$.
In mean-field spin glass models, $T_{\rm SEB}$ corresponds to the well-defined thermodynamic spin glass transition temperature $T_{\rm c}$, $T_{\rm SEB} = T_{\rm c}$. 
In finite-dimensional structural glasses, $T_{\rm SEB}$  corresponds to the mode-coupling theory (MCT) crossover temperature  $T_{\rm MCT}$, where the equilibrium relaxation time $\tau_{\rm eq}$ is expected to diverge. The difference between WEB and SEB is clearly visible in Fig.~\ref{fig:aging}c. The plateau (called a nonergodicity parameter $f$) associated with SEB appears in $F_{\rm eq}(t)$, which is the envelope of $F(t_{\rm w}, t_{\rm w} + t)$ curves. This equilibrium plateau is not the same plateau in $F_w(t_{\rm w})$ - they have different physical meanings as explained above. 

The main purpose of our theoretical modeling below is to explain the behavior of $C_w(t_{\rm w})$ and the difference between $T_{\rm WEB}$ and $T_{\rm SEB}$. We will begin by reviewing the  BTM model and discussing its inconsistency with simulation results. \\

\begin{figure*}[!htbp]
  \centering
\includegraphics[width=(0.7\linewidth)]{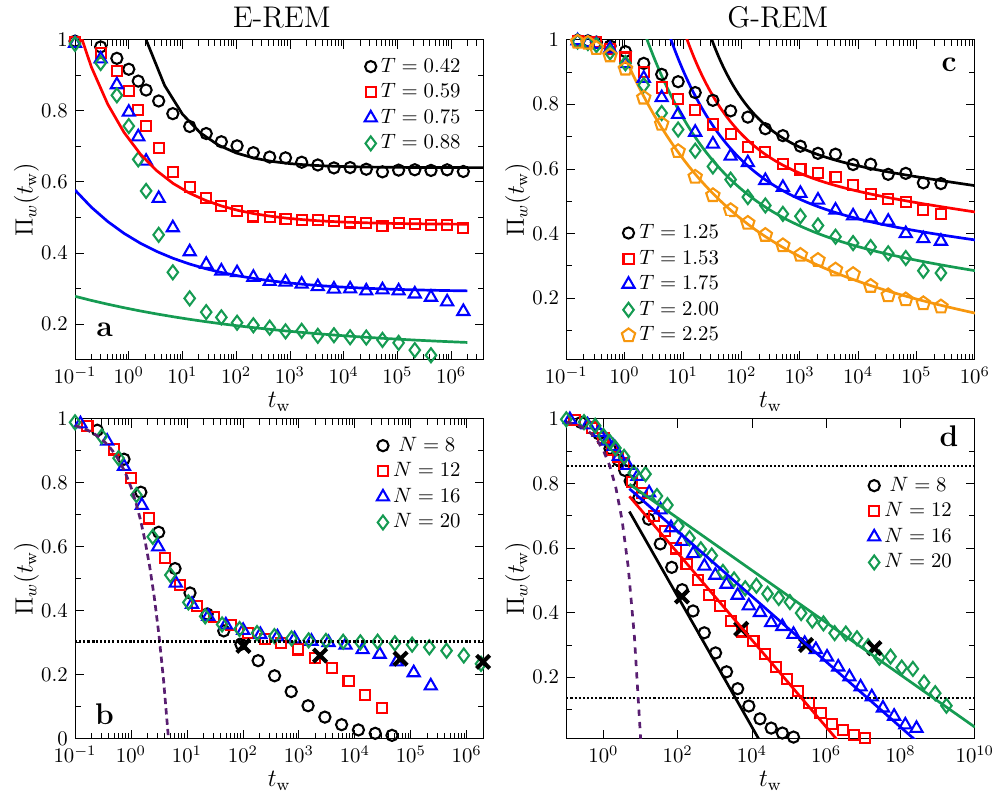}
  \caption{{\bf Aging function $\Pi_w(t_{\rm w})$ for the (a,b) E-REM and (c-d) G-REM.} 
  In both models, $w=0.5$.
  Open points are MC simulation data. 
(a) E-REM results with $N=20$  (points). Data in the plateau regime are fitted by $\Pi(t_{\rm w}) = H \left( 1 + A \, t_{\rm w}^{-\alpha} \right)$ (lines).  (b) E-REM data at $T=0.75$. The purple dashed line represents the short-time behavior $\Pi_w^{\rm s}(t_{\rm w}) = 1-Ct_{\rm w}$ with $C \approx 0.214$. The dotted horizontal line represents the BTM plateau $H(w,x) \approx 0.303$. The thermalization time $\tau_{\rm th} =  \exp(N/T)$ is marked by crosses.
(c) G-REM results with $N=128$ (points).  Data at large times are fitted by $\Pi(t_{\rm w}) = H \left( 1 + A \, t_{\rm w}^{-\alpha} \right) - k \ln t_{\rm w}$ (lines). 
(d) At $T=0.75$, the G-REM data are dominated by logarithmic decay $\Pi(t_{\rm w}) \sim - \frac{B}{N} \ln t_{\rm w}$ (lines) in small systems, where $B=0.7$ is a fitting parameter.
The  horizontal dashed lines represent the GTM plateau  $H(w,\hat{x}) \approx 0.854$ and the BTM plateau $H(w,x) \approx 0.136$.
The fitting parameters $H$ and $\alpha$ for both models are plotted in Fig.~\ref{fig:theory}(a,b).
  }
  \label{fig:rems}
\end{figure*}

\section{Bouchaud's trap model}
In the BTM, the trapping time $\tau$  in a metastable glass basin    is related to the energy barrier $\Delta E$ of that basin though the Arrhenius law, $\tau \sim  \exp(\Delta E/T)$. The trajectory of the random walker consists of a sequence of hops in the energy landscape with a trapping time distribution  $\psi(\tau)$ that is determined  by the energy barrier distribution $P(\Delta E)$. A further assumption is  that consecutive hops are uncorrelated (called {\it renewal mechanism}), i.e., $\psi(\tau_1, \tau_2) = \psi(\tau_1)\psi( \tau_2)$.

\begin{figure*}[!htbp]
    \includegraphics[width= 0.7\linewidth]{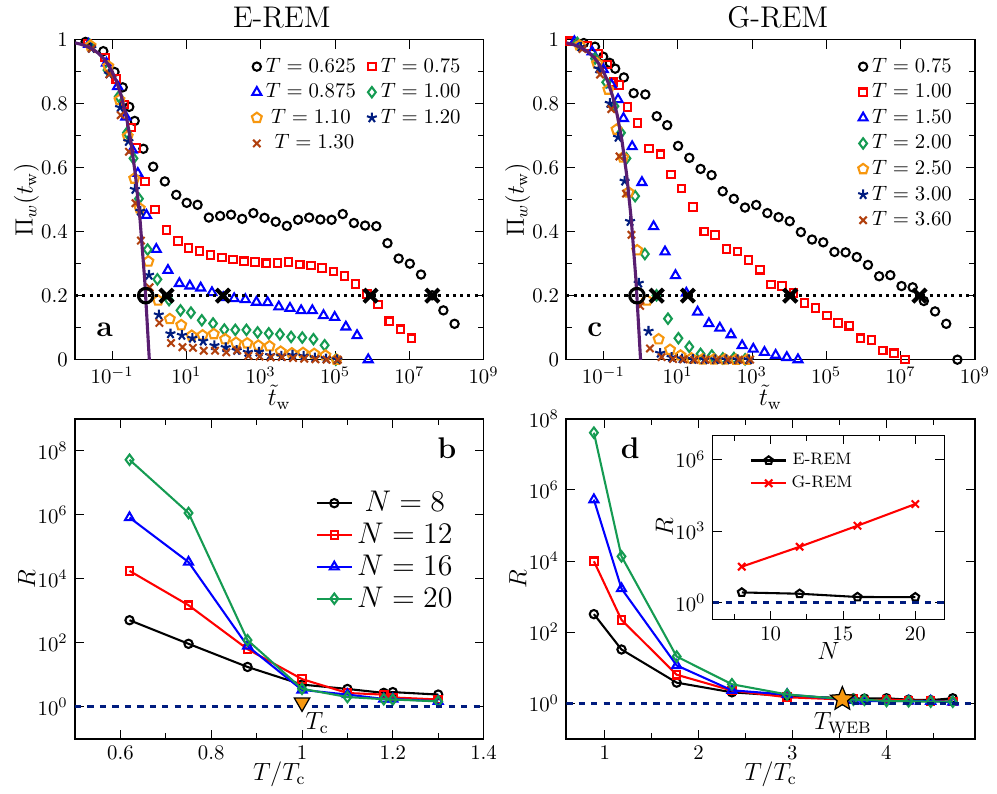}
\caption{{\bf WEB in (a,b) E-REM and (c,d) G-REM.}
In (a,c), $\Pi_w(t_{\rm w})$ is obtained with $N=20$ and $w=0.5$ by MC simulations at the rescaled time $\tilde{t}_{\rm w} = C t_{\rm w}$ with $C = \frac{w \hat{x}}{1+\hat{x}}$. The solid line is $\Pi^{0}_w(\tilde{t}_{\rm w})=1-\tilde{t}_{\rm w}$ (see Eq.~\ref{eq:scaling_GTM}). The dotted horizontal line is $\Pi_{\rm thr}=0.2$. The time scales $\tilde{\tau}^0$ (circle) and $\tilde{\tau}$ (cross) are marked. 
(b,d) $R=\tilde{\tau}/\tilde{\tau}^0$ as a function of  $T$ for  different $N$.
(inset) $R$ as a function of $N$ at $T=1.17 \, T_{\rm c}$.
}
\label{fig:DT}
\end{figure*}

The key input to the BTM is the energy barrier distribution $P(\Delta E)$ - once $P(\Delta E)$ is given, in principle the dynamics of the system can be reconstructed, with the  setup described above.
For example, one can compute $\Pi(t_{\rm w}, t_{\rm w} + t)$ that describes the probability of not leaving the glass basin between two times $t_{\rm w}$ and $t_{\rm w}+t$ after quenching, which is related to the spin auto-correlation function 
via $C^{\rm ag}(t_{\rm w}, t_{\rm w}+t) = q_{\rm EA} \Pi(t_{\rm w}, t_{\rm w} + t)$, with  $q_{\rm EA}$  the Edwards-Anderson order parameter
~\cite{bouchaud1998out}. 
The BTM  assumes an exponential distribution~\cite{bouchaud1992weak, bouchaud1995aging},
\beq
p_{\rm BTM}(\Delta E) \sim \exp\left(- {\Delta E}/T_{\rm c} \right).
\label{eq:BTM}
\eeq
In the REM, it was claimed that $T_{\rm c}$ 
corresponds to the spin glass transition temperature (see Appendix A for the details of the REM).  The exponential form Eq.~(\ref{eq:BTM}) is consistent with mean-field thermodynamic calculations for the glass models in the so-called  the one-step replica symmetry breaking (1-RSB) universality class~\cite{mezard1987spin}. Interestingly, Eq.~(\ref{eq:BTM}) is formally identical to the Boltzmann distribution,  $P(E) \sim \exp(-E/T)$, of the microstate energy in an equilibrium system -
the only difference is the replacement of $T$  by a constant $T_{\rm c}$.
Because $P(\Delta E)$ is a static property of the energy landscape, it should be $T$-independent; then a natural guess of the parameter $T_{\rm c}$  in Eq.~(\ref{eq:BTM}) is the glass transition temperature, which is the only characteristic temperature in simple models such as the REM.

According to Eq.~(\ref{eq:BTM}) and the Arrhenius law, $\tau$ follows a power-law distribution, $
\psi(\tau) \sim \tau^{-(1+x)}$, where $x = T/T_{\rm c}$   is the reduced temperature.
Because  $0<x<1$ below $T_{\rm c}$, the average trapping time  $\langle \tau \rangle$ diverges in the glass phase. Thus the system takes infinite time to reach equilibrium, leading to long-time aging and WEB. Mathematically, WEB is described by the
arcsin law~\cite{arous2006course} (see Appendix B):   
\beq
\begin{split}
\lim_{t_{\rm w} \to \infty}  \Pi_w(t_{\rm w}) 
= \frac{\sin(\pi x)}{\pi} \int_{w}^\infty \frac{du}{u^x(1+u)} 
 \equiv H(w,x).
\end{split}
\label{eq:Pi}
\eeq
Here $H(w,x)$ is a {\it weak ergodicity breaking order parameter}: $H(w,x)=0$ in the ergodic phase, and  $H(w,x)>0$ in the WEB phase. Importantly, the BTM predicts that the WEB and spin glass transition occurs simultaneously, $T_{\rm WEB} = T_{\rm c}$. We will show below that this prediction is inconsistent with simulation results.

With the given theoretical assumptions, the arcsin law Eq.~(\ref{eq:Pi}) is proven to be rigorous in the thermodynamic and large-time limits (taking $N \to \infty$ first and then $t_{\rm w} \to \infty$)~\cite{gayrard2016convergence, gayrard2019aging}. It was expected that such rigorous results would apply to the REM. However,  the arcsin law  is challenged by the single-spin flip MC simulations of the standard REM with a Gaussian distribution of configuration energy (G-REM), as shown in~\cite{baity2018activated}. Up to the largest system size and time ($N \sim 20$  and $t_{\rm w} \sim 10^{10}$) that can be simulated by regular CPUs, there is no sign that the simulation data of $\Pi_w(t_{\rm w})$ would converge to the predicted plateau $H(w,x)$  (see also Fig.~\ref{fig:rems}d). In contrast, such convergence is well observed in the REM  with an exponential distribution of energy (E-REM)~\cite{baity2018activated} (see also  Fig.~\ref{fig:rems}b). Thus the aging theory built on the   phenomenological BTM cannot fully explain the asymptotic dynamics in simple models such as the REM.

\begin{figure*}[!htbp]
    \includegraphics[width= \linewidth]{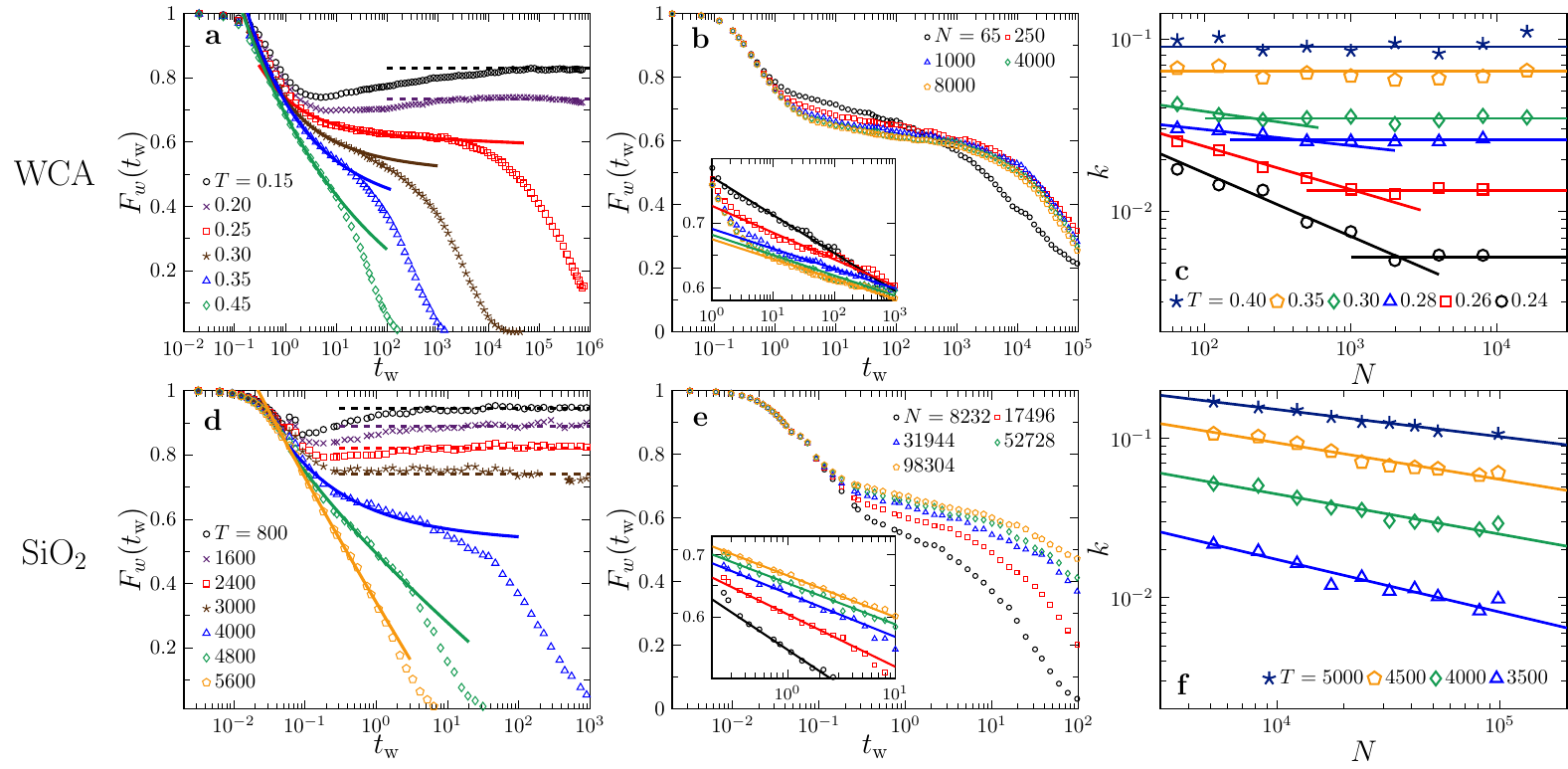}
\caption{{\bf Aging in  (a-c) WCA and (d-f) amorphous silica models.}
(a)  $F_w(t_{\rm w})$ obtained by MD simulations of the WCA model, with $N=16000$. The data in the intermediate time regime are fitted to $F_w(t_{\rm w}) = \tilde{H} \left( 1 + A \, t_{\rm w}^{-\alpha} \right) - k \ln t_{\rm w}$ (solid lines).
At the two lowest $T$, $\tilde{H}$ is estimated by fitting the plateau (dashed lines), and $\alpha$ cannot be determined due to the presence of a dip.
 (b) $F_w(t_{\rm w})$ at $T=0.26$ with different $N$. Inset is the close-up for the data in $1 < t_{\rm w} < 10^3$, fitted to $F_w(t_{\rm w}) \sim -k \ln t_{\rm w}$ (lines). 
The parameter $k$  obtained in this way is plotted in (c) as a function of $N$ at different $T$. The crossover size $n$ is determined by the intersect of two linear fitting lines (when $T<0.35$). 
(d-e) Similar plots for amorphous silica (${\rm SiO}_2$).
In (d,f), $N=31944$; in (e), $T=4000$ K.
In all panels, $w=1/2$. The parameters $H$ and $\alpha$ are plotted in Fig.~\ref{fig:theory}(a,b) for both models.
}
\label{fig:WCA}
\end{figure*}

\section{Generalized trap model} 
To understand why the BTM fails to describe the simulation results, one should naturally re-examine the underlying assumptions. Based on the numerical data obtained for REMs, we have examined three major assumptions as described above: (i) the Arrhenius law $\tau \sim  \exp(\Delta E/T)$, (ii) the renewal mechanism $\psi(\tau_1, \tau_2) = \psi(\tau_1)\psi( \tau_2)$, and (iii) the exponential barrier distribution Eq.~(\ref{eq:BTM}). We find that the first two assumptions hold (see Appendix A), but the third one requires correction (see below). Based on these observations, we  propose a GTM, which  retains  all assumptions of the original BTM,  except for the exponential barrier distribution. In the GTM, we add a Gaussian correction term to the barrier distribution,
\beq
p_{\rm GTM}(\Delta E) \sim \exp\left[- \frac{\Delta E}{T_{\rm c}} -\frac{(\Delta E - \bar{E})^2}{2Na}\right],
\label{eq:GTM}
\eeq
where $a$ is a model-dependent parameter. 
The energy scale $\bar{E} = b N$ is assumed to be extensive, where $b$ is also model-dependent.
Note that the Gaussian term, which follows the standard central limit theorem, is essentially caused by  finite-size effects. 
Equation~(\ref{eq:GTM}) will be explicitly examined in the REMs below, but for now, we take it as input and derive the corresponding aging function theoretically (see details in the Supplemental Material, SM, Sec. S1).

There are two crucial dynamical consequences due to the Gaussian term in Eq.~(\ref{eq:GTM}).
The first consequence is the modification of the asymptotic plateau $H(w,x)$ in Eq.~(\ref{eq:Pi}) with $x=T/T_{\rm c}$ 
replaced by an effective $\hat{x} = T/T_{\rm WEB}$. The second consequence introduces a logarithmic decay term $\sim \frac{1}{N}\ln t_{\rm w}$ in the pre-asymptotic behavior of $\Pi_w(t_{\rm w})$, which avoids the asymptotic plateau in small systems. Next we discuss them in detail.

Expanding Eq.~(\ref{eq:GTM}) gives (neglecting the constant term), $\ln p_{\rm GTM}(\Delta E)  \sim -\frac{1}{T_{\rm c}} \left(1-\frac{\bar{E}T_c}{Na} \right) \Delta E 
 - \frac{1}{2Na}\Delta E^2$.
 The  linear coefficient suggests an effective $x$ parameter,
$\hat{x} = \left(1- b T_{\rm c}/a \right)x = (1-W)T/T_{\rm c}$,
where the parameter $W = b T_{\rm c}/a$  weights the contributions of linear and Gaussian terms in Eq.~(\ref{eq:GTM}).
Importantly, $W$ is independent of $N$, and thus this modification does not disappear even in the thermodynamic limit. However, if we take $N \to \infty$ first 
in Eq.~(\ref{eq:GTM}), then obviously the GTM degenerates with  the BTM.
It suggests that the double limits $N \to \infty$ and $t_{\rm w} \to \infty$ are not interchangeable, once WEB occurs.  In the rigorous analyses performed previously~\cite{vcerny2017aging, gayrard2019aging, gayrard2019dynamic, derrida2023random}, the order is   $N \to \infty$ first and then $t_{\rm w} \to \infty$.
However, in MC simulations, generally long-time simulations ($t_{\rm w} \to \infty$) are performed for small $N$, then finite-size analysis is performed to extrapolate the behavior in the thermodynamic limit. In the latter case, the correction due to the Gaussian term is non-negligible even in large systems, which explains the discrepancy between simulation and rigorous results.

The asymptotic plateau of $H(w,x)$ still obeys the arcsin law  Eq.~(\ref{eq:Pi}), if $x$ is replaced by  $\hat{x}$.
Interestingly, $H(w,\hat{x})$ suggests that the WEB occurs at, 
\beq
T_{\rm WEB} = \frac{T_{\rm c}}{1 - W},
\label{eq:Td}
\eeq
which is above the glass transition temperature $T_{\rm c}$ .  The ratio $T_{\rm WEB}/T_{\rm c} = 1/(1-W)$
depends on the  model, since $W$ is model-dependent, as explained above.

The second interesting consequence is caused by the  quadratic term $\sim \Delta E^2/N$ in Eq~(\ref{eq:GTM}), which is $O(1/N)$. 
This term adds a logarithmic decay term (higher-order corrections are neglected) to the pre-asymptotic behavior of $\Pi_{w}(t_{\rm w})$, as {derived in SM Sec. S1}:
\beq
  \begin{split}
 & \Pi^{\rm GTM}_w(t_{\rm w})  \sim  \\ 
 &  
  \begin{cases}
 1- C \, t_{\rm w},  t_{\rm w} < \tau_{\rm m}, \\
 H \left( 1 + A \, t_{\rm w}^{-\alpha} \right) - \frac{B}{N} \ln t_{\rm w}, \tau_{\rm m} < t_{\rm w} < \tau_{\rm th}, \\
 {\rm thermalization}, t_{\rm w} > \tau_{\rm th}.\\
\end{cases}
\end{split}
\label{eq:scaling_GTM}
\eeq
Here $\tau_{\rm m} \sim O(1)$ is a microscopic time scale, and the rapid decay when $t_{\rm w} < \tau_{\rm m}$ is due to microscopic dynamics, with $C=\frac{w \hat{x}}{1+\hat{x}}$. The thermalization time $\tau_{\rm th}$ depends on the largest energy barrier $\Delta E_{\rm max}$ through $\tau_{\rm th} \sim \exp(\Delta E_{\rm max}/T)$; in the REMs, $\Delta E_{\rm max} \sim N$ and therefore $\tau_{\rm th} \sim \exp(N/T)$ (see Fig.~\ref{fig:rems}b,d).
We do not discuss the dynamics near and after $\tau_{\rm th}$ in this study, although thermalization itself is a very interesting problem~\cite{fermi1955studies}. 

The most important part in Eq.~(\ref{eq:scaling_GTM}) is the aging regime at intermediate times $\tau_{\rm m} < t_{\rm w} < \tau_{\rm th}$. Expressions of the coefficients are obtained by  theoretical calculations with the lowest-order approximation (see SM Sec.~S1):  $A   = \frac{1}{(1-\hat{x})\Gamma(1-\hat{x})\Gamma^2(\hat{x})}$ with $\Gamma(x)$ a gamma function, and $B= - \frac{T^2}{a} \frac{\partial H(w,\hat{x})}{\partial \hat{x}}$. The first term $ H \left( 1 + A \, t_{\rm w}^{-\alpha} \right)$  in Eq.~(\ref{eq:scaling_GTM}) tells us that, in the limit $N  \to \infty$, $\Pi_w(t_{\rm w})$ approaches the asymptotic limit 
$H(w,\hat{x})$ following a power-law function, with the exponent given by our theory:
\beq
\alpha = 1- \frac{T}{T_{\rm WEB}}.
\label{eq:alpha}
\eeq
The second term  $- \frac{B}{N}  \ln t_{\rm w}$ implies that, when $N$ is small, this logarithmic decay will destroy the asymptotic plateau - this is what we observe in simulations of G-REM.  

It can be shown that the theoretical predictions from the GTM capture universal aging behavior in simulated spin and structural glasses: (i) with the model-dependent temperature $T_{\rm WEB}$ determined, the WEB order parameter $H$ and the exponent $\alpha$ estimated from the simulation data collapse when plotted according to  Eqs.~(\ref{eq:Pi}) and (\ref{eq:alpha}); (ii) the logarithmic decay $- k  \ln t_{\rm w}$ in $\Pi_w(t_{\rm w})$ can be universally observed at appropriate time and temperature windows in different models, and the coefficient $k$ can be used to estimate the size of activation clusters. Below we demonstrate such universality with simulations of several spin and structural glass models. \\

\begin{figure}[!htbp]
    \includegraphics[width= 0.9\linewidth]{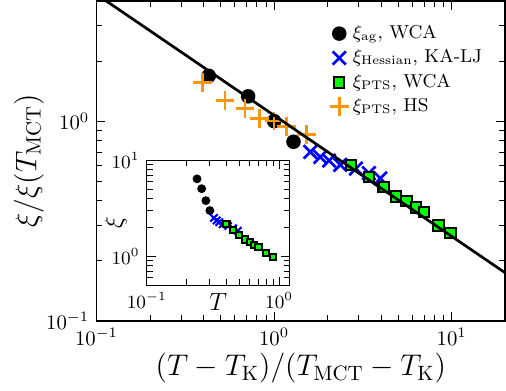}
\caption{{\bf Static lengths.}
The following data are plotted:  $\xi_{\rm ag}$ for the WCA model (this work), $\xi_{\rm Hessian}$ for the KA-LJ model from~\cite{karmakar2012direct} ($T$ is rescaled by a factor of $3/4$), $\xi_{\rm PTS}$ for the WCA model from~\cite{hocky2012growing}, and $\xi_{\rm PTS}$ for HSs from ~\cite{berthier2017configurational}.
The HS $\xi_{\rm PTS}$  is plotted as a function of $(Z^{-1}-Z^{-1}_{\rm K})/(Z_{\rm MCT}^{-1}-Z^{-1}_{\rm K})$.
The black line represents the fitting  $\xi \sim (T-T_{\rm K})^{-\nu}$, where $T_{\rm K}=0.21$ and $\nu = 0.61(1).$ (inset) $\xi$ as a function of $T$.
}
\label{fig:length}
\end{figure}

\begin{figure*}[!htbp]
  \centering 
  \includegraphics[width=0.8\linewidth]{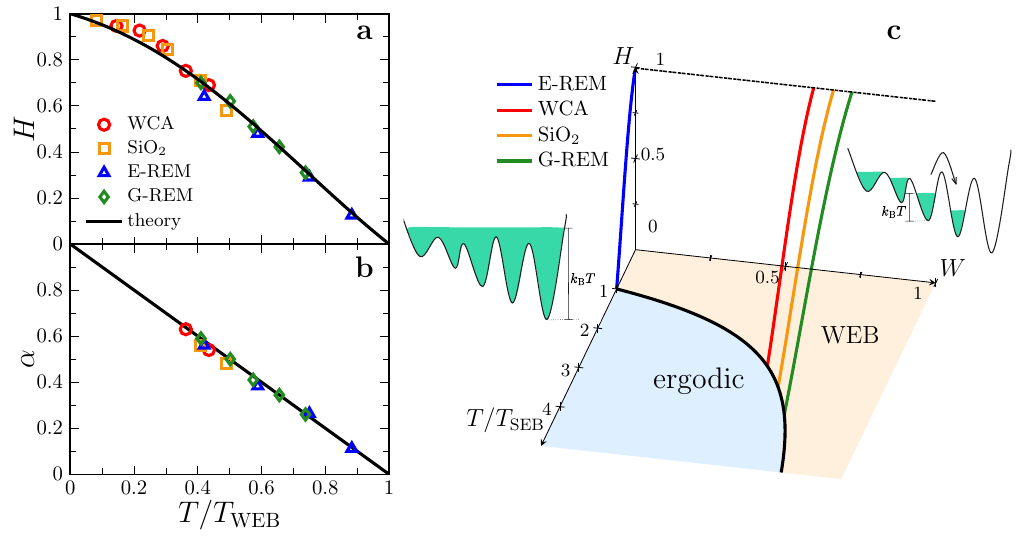}
  \caption{{\bf Unified phase diagram of activated aging and WEB in glasses.}
  (a) WEB order parameter $H$ and (b) power-law convergence exponent $\alpha$ obtained by simulations of four glass models ($w=t/t_{\rm w} = 1/2$). The theoretical lines are Eqs.~(\ref{eq:Pi}) and~(\ref{eq:alpha}). 
 (c) In the unified phase diagram, WEB and ergodic phases are separated by the $T/T_{\rm SEB} = 1/(1-W)$ line, and the $H$-axis quantifies the order parameter. Each glass model corresponds to one line in the plot with the corresponding  $W$.  
 Insets are schematic energy landscapes in ergodic and WEB phases; the arrow indicates hopping to lower energy basins during aging.}
  \label{fig:theory}
\end{figure*}

\section{Aging in spin glass random energy models}
In REMs, we examine both the assumptions and predictions of the GTM.
We first focus on the assumptions. 
As shown in Appendix A, data obtained from numerical simulations are consistent with the Arrhenius law and renewal mechanism.
Next we examine the barrier energy distribution,
Eq.~(\ref{eq:GTM}), using two independent approaches. The first one is an exhaustive  enumeration of energy barriers using the barrier-tree (BT) algorithm~\cite{flamm2000rna, ferreira2000landscape, fontanari2002fractal, flamm2002barrier} (see Fig.~\ref{fig:barrier}a for an example of a sub-tree and Appendix~A for algorithm details). The BT method  provides the exact distribution $p^{\rm BT}(\Delta E)$, but is restricted to small systems ($N \leq 20$). We also develop a tree-expansion (TE) theory to compute the distribution analytically and approximately, giving  $p^{\rm TE}(\Delta E)$  for any $N$ ({see SM Sec. S2}). The difference between $p^{\rm BT}(\Delta E)$ and  $p^{\rm TE}(\Delta E)$ is unnoticeable, as shown in
Fig~\ref{fig:barrier}(b-d).

While the distribution $p_{\rm E-REM}^{\rm BT}(\Delta E)$ of the E-REM  obtained by the BT algorithm
 is very close to the exponential form $p_{\rm BTM}(\Delta E)$ (see Eq.~\ref{eq:BTM}),  it is evident that the G-REM distribution $p_{\rm G-REM}^{\rm BT}(\Delta E)$ deviates from a pure exponential distribution (Fig.~\ref{fig:barrier}b).  To analyze this deviation more carefully, the exponential part Eq.~(\ref{eq:BTM}) is subtracted from the full distribution, and the remainder $\ln [ p^{\rm BT}(\Delta E)] + \Delta E/T_{\rm c}$ is plotted in
Fig.~\ref{fig:barrier}c for both models. The data for G-REM in
Fig.~\ref{fig:barrier}c can be well fitted by a quadratic function $f(\Delta E) = c -\frac{(\Delta E - \bar{E})^2}{2\sigma^2}$. The variance $\sigma^2 = a N $ and mean $\bar{E} = b N$ linearly depend on $N$, as shown in Fig.~\ref{fig:barrier}d. From the linear fitting, we obtain {$a=1.12$ and $b = 0.96$} for the G-REM. In contrast, the remainder of the E-REM in
Fig.~\ref{fig:barrier}c is negligible. Thus for the E-REM, the barrier distribution is well described by a pure exponential function as in the BTM.

Next we compare the GTM prediction Eq.~(\ref{eq:scaling_GTM})  with the MC simulation data of  REMs. For the E-REM, because the correction term in Eq.~(\ref{eq:GTM}) is negligible, we can set $T_{\rm WEB} = 
 T_{\rm c} = 1$ and expect  $\Pi_w(t_{\rm w}) \sim H \left( 1 + A \, t_{\rm w}^{-\alpha} \right)$ in the aging regime. 
Fitting the data by this form, we obtain $H$  and $\alpha$ for a few $T$ with a fixed system size
$N=20$ (Fig.~\ref{fig:rems}a). The fitted values of $H$  and $\alpha$ are compared to the theoretical predictions Eq.~(\ref{eq:Pi}) and~(\ref{eq:alpha}) in Fig.~\ref{fig:theory}(a,b), and good agreement is found.
In Fig.~\ref{fig:rems}b, we examine the finite-size effects, at a fixed $T=0.75$. It can be seen that even for small systems $N=8-20$, the BTM plateau can be well-observed. The analysis confirms that the logarithmic decay in Eq.~(\ref{eq:GTM}) is not essential in E-REM, as expected.

For the G-REM, the Gaussian correction in Eq.~(\ref{eq:GTM})
is non-negligible as shown in Fig.~\ref{fig:barrier}. Consequently, we need to use the full expression $\Pi_w(t_{\rm w}) \sim H \left( 1 + A \, t_{\rm w}^{-\alpha} \right) - k \ln t_{\rm w}$ to fit the data (see Fig.~\ref{fig:rems}c). 
Here we have set $\hat{x} = T/T_{\rm WEB}$, $T_{\rm WEB} = T_{\rm c}/(1-W) \approx 3.6 \, T_{\rm c}$
(see Eq.~\ref{eq:Td}), where $W=bT_{\rm c}/a \approx 0.722$  with $a = 1.12$ and $b = 0.96$ from Fig.~\ref{fig:barrier} and $T_{\rm c} \approx 0.849$ from the literature~\cite{derrida1980random}.
The fitted $H$ and $\alpha$ are compared to the theory in Fig.~\ref{fig:theory}(a,b). The coefficient $k$ depends on $N$ as expected by the theory $k \sim 1/N$ (see Fig.~\ref{fig:rems}d).

An interesting prediction from the GTM theory is that the WEB does not necessarily co-occur with  the spin glass transition: for the E-REM, $T_{\rm WEB} = T_{\rm c}$, while for the G-REM, $T_{\rm WEB} \approx 3.6 \, T_{\rm c}$. 
The factor 3.6 in the G-REM is significantly larger than one - thus the GTM scenario ($T_{\rm WEB} \approx 3.6 \, T_{\rm c}$) can be unambiguously tested against the original BTM scenario ($T_{\rm WEB} = T_{\rm c}$)~\cite{bouchaud1992weak}.
Below we directly verify the GTM scenario. 
In Fig.~\ref{fig:DT} we plot  the MC data of $\Pi_w(t_{\rm w})$ for both E-REM and G-REM, with $t_{\rm w}$ rescaled by $C=\frac{w\hat{x}}{1+\hat{x}}$ to collapse the short-time behavior $\Pi_w^{\rm 0}(t_{\rm w}) = 1-C t_{\rm w}$ at different $T$ (note that $C$ is not a fitting parameter). 
If WEB occurs, $\Pi_w(\tilde{t}_{\rm w})$ should not vanish in the $\tilde{t}_{\rm w} \to \infty$ limit, where $\tilde{t}_{\rm w} \equiv C \,{t}_{\rm w}$ is the reduced waiting time.
We define a time scale $\tilde{\tau}$ such that $\Pi_w(\tilde{t}_{\rm w} = \tilde{\tau}) =  \Pi_{\rm thr}$ and choose a threshold $\Pi_{\rm thr} = 0.2$. 
At sufficiently high $T$ (in the ergodic phase), $\tilde{\tau}= \tilde{\tau}^0$ is a constant, where $\tilde{\tau}^0$ is defined by $\Pi_w^0(\tilde{t}_{\rm w} = 
\tilde{\tau}^0) = 1- \tilde{\tau}^0 = \Pi_{\rm thr}$.
As shown in Fig.~\ref{fig:DT}, with an increasing $T$, the ratio $R = \tilde{\tau}/\tilde{\tau}^0$ approaches one around $T_{\rm c}$ in E-REM and around $T_{\rm WEB} \approx 3.6 \, T_{\rm c}$ in G-REM. According to the GTM scenario, we expect $R \to \infty$ below $T_{\rm WEB}$ in the thermodynamic limit, because $\tilde{\tau}$ should diverge in the WEB phase. This expectation is confirmed by finite-size analyses at a fixed $T=1.17\,T_{\rm c}$ slightly above $T_{\rm c}$ in both models (see Fig.~\ref{fig:DT}d-inset). In G-REM, $R$ increases with $N$ at this temperature; in contrast, in E-REM, $R \approx 1$ is nearly invariant. This is direct evidence of WEB above $T_{\rm c}$  in the G-REM. \\

\begin{table*}[!htbp]
\centering
\caption{{\bf Characteristic temperatures}. 
The spin glass transition temperature $T_{\rm c}$ is obtained by thermodynamic calculations for G-REM~\cite{derrida1980random} and E-REM~\cite{bouchaud1997universality}.
The divergence temperature  $T_{\rm 0}$ is estimated from the VFT fitting of the  $\alpha$-relaxation time $\tau_{\rm eq} \sim \exp\left(\frac{C}{T-T_{\rm 0}}\right)$ for the WCA model{~\cite{probing2014wang}}.
The MCT temperature $T_{\rm MCT}$ is estimated from the MCT fitting, $\tau_{\rm eq} \sim (T-T_{\rm MCT})^{-\gamma}$ for the WCA model~\cite{critical2010berthier}, and $D \sim (T-T_{\rm MCT})^{\gamma}$ of 
the diffusion constant $D$ for amorphous $\rm{SiO}_2$~\cite{static1999horbach}.
The onset temperature $T_{\rm onset}$ in the WCA model corresponds to
the crossover temperature where the Stokes-Einstein relation, $\tau_{\rm eq} \sim D^{-1}$, breaks down~\cite{role2011berthier}.
Using four different criteria, Ref.~\cite{banerjee2017determination} reports 
$T_{\rm onset} = 0.6-0.7$ for the WCA model.
Data without references are obtained by the current study (see the main text and Appendix for details).
}
\begin{tabular}{ l  l l  l  l l l l}
\hline
\hline
& $T_{\rm c}$ & $T_{\rm 0}$& $T_{\rm MCT}$ & $T_{\rm onset}$ &   $T_{\rm K}$& $T_{\rm WEB}$ & W\\
\hline
G-REM & $\frac{1}{\sqrt{2 \log 2}}${\cite{derrida1980random}} & - & - & - & - & 3.05 & 0.722 \\
\hline
E-REM & 1~\cite{bouchaud1997universality} & - &  - & - & - & 1 & 1\\
\hline
WCA & - & 
 0.158~\cite{probing2014wang} 
& 0.28~\cite{critical2010berthier} &
 \begin{tabular}{@{}l@{}} 0.513~\cite{role2011berthier} \\ 0.6-0.7~\cite{banerjee2017determination}\end{tabular}
& 0.21 & 0.69 & 0.594\\  
\hline
$\rm{SiO}_2$ & - &  - 
& $3330 {\rm K}$~\cite{static1999horbach} & 5800{\rm K} 
& - & {9800}{\rm K} & 0.66\\
\hline 
\hline
\end{tabular}
\label{table:T}
\end{table*}

\section{Aging in structural glasses}
To further examine the universality of the GTM theory, we simulate two structural glass formers: the WCA model as an example of simple molecular systems, and the amorphous silica ($\rm{SiO}_2$) model as an example of network glasses (see Appendixes~C and D for details). 
Because efficient sampling of energy barriers in structural glasses remains technically challenging~\cite{denny_trap_2002, heuer2008exploring, ji2025role}, we will only compare the simulated aging function $\Pi_w(t_{\rm w})$ to theoretical predictions (see the data in Fig.~\ref{fig:WCA} with a fixed $w=1/2$). Note that $\Pi_w(t_{\rm w})$ is related to $F_w(t_{\rm w})$ via $F_w(t_{\rm w}) =  f \Pi_w(t_{\rm w})$, where $f$ is the nonergodicity parameter defined by the large-time plateau value of the equilibrium  incoherent scattering function $f = \lim_{t \to \infty}F_{\rm eq}(t)$.
The nonergodicity parameter $f$ is analogous to the Edwards-Anderson order parameter $q_{\rm EA}$ in spin glasses, both of which are determined by the typical size of the glass basins.   

We first extract $H$ and $\alpha$ from the aging data, and compare them to theoretical predictions. 
As shown in Fig.~\ref{fig:WCA}(a,d),  the high-$T$ simulation results can be fitted to the form, {$F_w(t_{\rm w}) = \tilde{H} \left( 1 + A \, t_{\rm w}^{-\alpha} \right) - k \ln t_{\rm w}$, from which $\tilde{H}$ and $\alpha$ are obtained. Note that this expression only works in the intermediate-$t_{\rm w}$ regime (see Eq.~\ref{eq:scaling_GTM}), and thus the small-$t_{\rm w}$ (microscopic dynamics) and large-$t_{\rm w}$ (thermalization) data points should be excluded from the fitting.
At lower $T$, a dip in $F_w(t_{\rm w})$, which might be attributed to the so-called boson peak~\cite{habasaki1995origins}, appears after the short-time processes, making the above fitting form inapplicable. Nevertheless, at these temperatures, $\tilde{H}$ can still be estimated since the plateau is well-defined. 
Then the relation $H = \tilde{H}/f$ is used to estimate $H$ (see Appendix E for the data of $f$).}
The values of $H$ and $\alpha$ are plotted in Fig.~\ref{fig:theory}(a,b), as functions of $\hat{x}=T/T_{\rm WEB}$ with a fixed $w=1/2$.  With a proper choice of $T_{\rm WEB}$, these data collapse with the GTM theoretical curves Eq.~(\ref{eq:Pi}) and~(\ref{eq:alpha}). Note that $T_{\rm WEB}$ is obtained from fitting the data to the arcsin law Eq.~(\ref{eq:Pi}). 
A further consistent check is illustrated in Fig.~\ref{fig:aging}d for the WCA model: using the same $T_{\rm WEB}$ and $f$, the theoretical curve $\tilde{H}(w, \hat{x}=0.36) =  f H(w, \hat{x}=0.36)$, with $H$ given by Eq.~(\ref{eq:Pi}), agrees with the simulation data $F^{\rm ag}(w)$. Together, Fig.~\ref{fig:aging}d and~\ref{fig:theory}(a,b) suggest that the theory works for different $T$ and $w=t/t_{\rm w}$.

The finite-size dependence of the coefficient $k$ in the logarithmic term $- k \ln t_{\rm w}$ reveals very interesting properties of activated processes.  According to Eq.~(\ref{eq:scaling_GTM}), we expect $k \sim 1/N$ in mean-field models. This $N$ essentially appears in the Gaussian term in Eq.~(\ref{eq:GTM}). 
In non-mean-field systems such as the WCA and  amorphous silica  models, $N$ should be replaced by the size $n$ of clusters  involved in the activation events, because $n$ determines the true scale of the energy barrier. Such clusters are also called ``mosaics'' in the  RFOT~\cite{kirkpatrick1989scaling}.
If the system size $N$ is smaller than the characteristic size $n$, then all particles participate in  the activated aging, and $k(N)$ would depend on $N$. On the other hand, if $N > n$,  $k(N)$ would be independent of $N$ because $n$ imposes a cutoff of the relevant size. In other words, the finite-size data $k(N)$ can provide an estimate of the characteristic size $n$: with an increasing $N$, the data of $k(N)$ should reach a constant at $N=n$. Furthermore, we can obtain a length scale  $\xi_{\rm ag} \sim n^{1/d}$. Note that $\xi_{\rm ag}$ is a static length scale because it is time-independent. This length essentially reflects properties of the energy landscape, but remarkably, it is extracted from aging dynamics.

Based on the above consideration, we focus on the logarithmic part of the  $F_w(t_{\rm w})$ data, and perform  fitting using the form $F_w(t_{\rm w}) \sim - k \ln t_{\rm w}$
(see Fig.~\ref{fig:WCA}b,e) to obtain $k(N)$ (see Fig.~\ref{fig:WCA}c,f). In the WCA model, $k \sim N^{-\beta}$ with $\beta \approx 0.1-0.4$ when $N<n$ (see Appendix F), and $k \sim {\rm constant}$ when $N>n$. From these data, the characteristic size $n$ and length $\xi_{\rm ag} \equiv \frac{1}{2}(n/\rho)^{1/3}$ are determined. Note that there is no adjustable parameter in the definition of $\xi_{\rm ag}$  - the coefficient $1/2$ follows the convention that the static length is defined as a half of the box length. 
In the amorphous silica model, the plateau in $k(N)$ is not observed within the largest simulated system size. This difference is very interesting, because it highlights the distinct nature of the activation energy in the two models. 
In large  WCA systems ($N>n$),  the barrier energy $\Delta E[n(T)]$ is determined by the $T$-dependent $n(T)$ (note that $n(T)$ grows with a decreasing $T$), but not the system size $N$. In contrast, in the amorphous silica model $\Delta E(N)$ depends only on $N$, which suggests that $\Delta E$ is a   $T$-independent constant for a given system size. Correspondingly, the equilibrium relaxation time  behaves as $\tau_{\rm eq} \sim \exp[\Delta E(T)/T]$ and $\tau_{\rm eq} \sim \exp(\Delta E/T)$ in the two models - this is consistent with the well-known fact that WCA and amorphous silica  belong  to fragile and strong glass formers respectively~\cite{berthier2011theoreti} (see Fig.~\ref{fig:T_onset_silica}a). \\

\section{Growth of the static length in the WCA model}
In Fig.~\ref{fig:length}, the characteristic length $\xi_{\rm ag}$ in the WCA model is compared to two other static lengths. The first one is the PTS length $\xi_{\rm PTS}$ that characterizes the influence of fixed boundaries on the ergodicity of the bulk particles in the cavity~\cite{bouchaud2004adam}. The $\xi_{\rm PTS}$ data plotted in Fig.~\ref{fig:length} are obtained from Ref.~\cite{hocky2012growing} for the WCA model at the same density $\rho = 1.2$. The second one is the length scale $\xi_{\rm Hessian}$ used to collapse the data of the smallest eigenvalue of the Hessian matrix at different $T$~\cite{karmakar2012direct}, which matches $\xi_{\rm PTS}$ as noticed in Ref.~\cite{karmakar2015length}. The length $\xi_{\rm Hessian}$ has been measured for the Kob-Anderson  binary Lennard-Jones (KA-LJ) model~\cite{karmakar2012direct} but not yet for the WCA model. Ref.~\cite{hocky2012growing} shows that the PTS lengths, $\xi_{\rm PTS}(T)$, for the two models (WCA and KA-LJ) collapse if $T$ is rescaled by a factor of $4/3$, $\xi_{\rm PTS}^{\rm WCA}(T) = \xi_{\rm PTS}^{\rm KA-LJ}(4T/3)$. Based on this observation, in Fig.~\ref{fig:length} we plot $\xi_{\rm Hessian}$ of the KA-LJ model from ~\cite{karmakar2012direct} as a function of $3T/4$. 
Surprisingly, we find that not only $\xi_{\rm Hessian}$ and $\xi_{\rm PTS}$ collapse on the same curve as expected~\cite{karmakar2015length}, the length  $\xi_{\rm ag}$ also collapses. The growth range of each single length is limited within a factor of $2\sim3$, due to different numerical challenges: to obtain  $\xi_{\rm Hessian}$ and $\xi_{\rm PTS}$ one needs to equilibrate the system, which is impossible at a low $T$; to obtain $\xi_{\rm ag}$, one needs to fit the logarithmic decay of the aging function (see Fig.~\ref{fig:WCA}), which becomes difficult at a high $T$ due to rapid thermalization. However, if we combine the three lengths, the variation covers one decade. 

We can now  perform a scaling analysis of the combined $\xi$ data of  $\xi_{\rm Hessian}$, $\xi_{\rm PTS}$ and $\xi_{\rm ag}$. According to the RFOT, $\xi$ diverges at the Kauzmann temperature (ideal glass transition temperature) $T_{\rm K}$, as $\xi \sim (T-T_{\rm K})^{-\nu}$, where $\nu = 1/(d-\theta) = 2/d = 2/3$, with $\theta = d/2$  the surface tension exponent and $d=3$ the dimensionality~\cite{kirkpatrick1989scaling, lubchenko2007theory}. The best fitting of the data in Fig.~\ref{fig:length} gives $T_{\rm K} \approx 0.21$ and $\nu = 0.61(1)$. The estimated Kauzmann temperature $T_{\rm K} \approx 0.21$ is above, but not far away from, the reported divergence temperature $T_0 = 0.158$ of the relaxation time $\tau_{\rm eq} \sim \exp (\frac{C_0}{T-T_0})$ obtained by the Vogel–Fulcher–Tamann (VFT) fitting~\cite{probing2014wang}. The exponent $\nu = 0.61(1)$ is consistent with the value $\nu = 2/3$ given by the theoretical argument~\cite{kirkpatrick1989scaling, lubchenko2007theory}. The data do not support the possibility of diverging $\xi(T)$   at $T_{\rm K} = 0$ following a power-law (Fig.~\ref{fig:length}-inset).

In a recent study, the PTS length $\xi_{\rm PTS}$ is estimated for hard spheres (HSs) in 
an unprecedentedly supercooled regime by utilizing an efficient  swap MC algorithm, with the Kauzmann point estimated at $Z_{\rm K} \approx 45$ by extrapolating  the vanishing point of the configurational entropy~\cite{berthier2017configurational}. An analogy has also been  suggested  between $1/T$ in molecular systems and the reduced pressure (also called compressibility factor) $Z = P/(\rho T)$ in HSs, by showing the data collapse of the relaxation time as a function of $1/T$ in the former and $Z$ in the latter~\cite{dauchot2023glass, ortlieb2023probing}. Inspired by this analogy, in Fig.~\ref{fig:length}, we include the data of $\xi_{\rm PTS}$ for HSs. Very interestingly, if $\xi$ is rescaled by the value $\xi(T_{\rm MCT})$ at the MCT crossover, and then plotted as a function of $(T-T_{\rm K})/(T_{\rm MCT}-T_{\rm K})$ for the WCA model, and of $(Z^{-1}-Z_{\rm K}^{-1})/(Z_{\rm MCT}^{-1}-Z_{\rm K}^{-1})$ for the HS model, all data collapse on the same curve. The additional data from HSs thus strengthen the above analysis following the RFOT scaling. 
\\

\section{Universal behavior and a unified phase diagram of aging} 
The above aging data of spin and structural glasses can be universally described by the same theoretical framework. The WEB temperature $T_{\rm WEB}$ is a model-dependent parameter (see Table~\ref{table:T}). With a given $w$, the WEB order parameter $H$ and exponent $\alpha$ in Eq.~(\ref{eq:scaling_GTM}), obtained from four different models, collapse onto the same theoretical curves given by Eq.~(\ref{eq:Pi}) and (\ref{eq:alpha}), as functions of $\hat{x} = T/T_{\rm WEB}$ (see Fig.~\ref{fig:theory}a,b).

Figure~\ref{fig:theory}c is a unified phase diagram for the aging behavior in spin and structural glasses.
The ergodicity in equilibrium dynamics breaks down at 
$T_{\rm SEB}$ ($T_{\rm SEB} = T_{\rm c}$ in REMs, and $T_{\rm SEB} = T_{\rm MCT}$ in structural glasses). 
The ratio between $T_{\rm WEB}$  and $T_{\rm SEB}$ is specified by  the model-dependent parameter $W$, $T_{\rm WEB}/T_{\rm SEB} = 1/(1-W)$ (see Eq.~\ref{eq:Td}). Each model corresponds to a line in Fig.~\ref{fig:theory}c with a fixed $W$; the WEB occurs at $T=T_{\rm WEB}$, where the order parameter vanishes, $H = 0$. The boundary between ergodic and WEB phases in the $T/T_{\rm SEB} - W$ plane is defined by the line of $T/T_{\rm SEB} = 1/(1-W)$ (i.e., $T=T_{\rm WEB}$). In Table~\ref{table:T}, the values of relevant parameters are summarized for the four models studied; among the listed characteristic temperatures, $T_{\rm WEB}$ appears to be the highest one.  
We expect that this phase diagram can generally include many other glass systems.\\

\section{Discussion} 
In the present work, aging dynamics of several spin and structural glasses are studied under a unified framework 
built on the GTM. According to the replica theory~\cite{mezard1987spin}, these models belong to the universality class of 1-RSB. In the future, it will be interesting to generalize the present approach to glass systems that have a hierarchical energy landscape (full-step replica symmetry breaking universality class)~\cite{bouchaud1995aging}, such as the Sherrington-Kirkpatrick spin glass model~\cite{sherrington1975solvable} and hard sphere glasses in the Gardner phase~\cite{charbonneau2014fractal, berthier2016growing, urbani2023gardner}.

In this study, we focus on the temperature regime where activation governs aging dynamics. Our preliminary simulation results of structural glasses suggest that,  at very low temperatures, the activation-dominated logarithmic energy decay is switched to a mean-field-like power-law decay~\cite{folena2020rethinking, nishikawa2022relaxation}.
A complete picture should be based on a more systematic exploration of  the competition between  activated and  mean-field aging~\cite{carbone2022competition}.

As a phenomenological model, the landscape-based GTM is unnecessarily incompatible with other mechanisms based on microscopic properties or processes. For example, the origin of the barrier energy distribution $P(\Delta E)$ could be related to the pinning energy of domain walls~\cite{nattermann1988random}, the energy required to flip a domain of spins in kinetically constrained models~\cite{sollich1999glassy}, or the activation energy for local rearrangements of particles~\cite{ji2025role}. How to reconcile the current landscape interpretation with microscopic mechanisms will be left for future investigations.

The discussed aging phenomenon is relevant  to many other disordered systems.
The two-time correlation functions of metallic glasses  measured  by the X-ray photon correlation spectroscopy~\cite{zhang2023pressure} can be analyzed to compare with our theoretical predictions. Studies on the non-equilibrium inter-domain dynamics of single protein molecules 
report a power-law dependence of the  characteristic relaxation time $\tau_{\rm c}$ on the observation time $t$, $\tau_{\rm c} \sim t^{0.9}$~\cite{hu2016dynamics}, very close to the linear relationship $\tau_{\rm neq} \sim t_{\rm w}$ observed in Fig.~\ref{fig:aging}b.
Finally, in the training dynamics of deep learning, the logarithmic decay of energy (loss function), similar to Fig.~\ref{fig:aging}a, is found to be responsible for the gain of generalization ability, 
in a recently developed thermal deep learning machine~\cite{huang2025liquid}.
Generalization of the current approach to these systems is expected in the future.

\begin{acknowledgments}
We warmly thank Wenyue Fan, Gang Huang, Wencheng Ji, Walter Kob, Jian Liu, C. Patrick Royall, Yujie Wang, Hajime Yoshino and Haijun Zhou for discussions. 
We acknowledge financial support from NSFC (Grants 12161141007, 11935002, 12047503, and 12404290), from Chinese Academy of Sciences (Grant ZDBS-LY-7017 and KGFZD-145-22-13) and from Wenzhou Institute (Grant WIU- CASQD2023009).
In this work access was granted to the High-Performance Computing Cluster of Institute of Theoretical Physics - the Chinese Academy of Sciences.\\
\end{acknowledgments}

\bigskip

\centerline{\Large \bf Appendix}

\section*{Appendix A: Random energy models}

\subsection*{1. Model}

A REM comprises $2^N$ configurations. Each configuration consists of $N$ Ising spins, whose energy is drawn randomly from a Gaussian probability distribution,
\begin{equation}
\rho_{\rm Gauss}(E) = \frac{1}{\sqrt{2 \pi N}} \exp\left (- E^2/2N \right).
\end{equation}
This original version with the Gaussian distribution $\rho_{\rm Gauss}(E)$ is called a Gaussian random energy model (G-REM). An alternative version, called an exponential random energy model (E-REM)~\cite{bouchaud1997universality, baity2018activated}, has been introduced previously, with  $\rho_{\rm Gauss}(E)$ replaced by 
\begin{equation}
\rho_{\rm exp}(E)=\frac{1}{T_{\rm c}} \exp(E/T_{\rm c})\Theta(-E),
\end{equation}
where $\Theta(x)$ is the Heaviside step function. 
In each realization, the energy assignment of $2^N$  configurations is fixed (quenched disorder)
in the following  static barrier-tree analyses and dynamical simulations. The procedure is then repeated for {$\sim 1000-20000$} realizations  to take the statistical average.

\begin{figure}
    \centering
    \includegraphics[width=\linewidth]{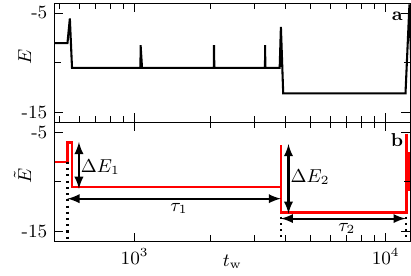}
    \includegraphics[width=\linewidth]{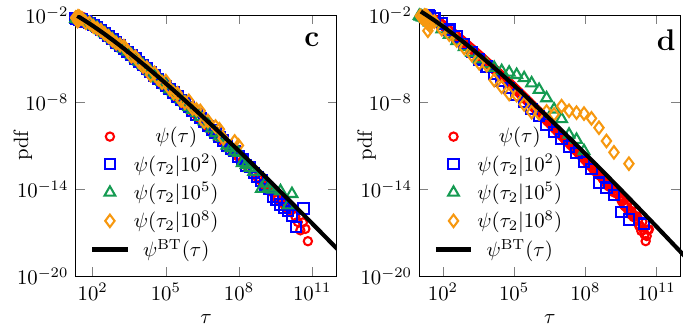}
    \caption{
    {\bf Basin hopping dynamics in MC simulations.}
    Data are obtained for the G-REM at $T=0.75$ with $N=20$.
    (a) An example of configuration  trajectory, where $E(t_{\rm w})$ is the energy of the configuration at $t_{\rm w}$. (b) Corresponding basin trajectory, where $\tilde{E}(t_{\rm w})$ is the energy of the basin or the saddle point at $t_{\rm w}$. With this, a sequence of hopping events with the barrier energy $\Delta E_{i_k}$ and the hopping time $\tau_k$ are identified. 
    (c) The hopping time probability distribution function (pdf) $\psi(\tau)$  and the conditional distribution  $\psi(\tau_2|\tau_1)$ measured in MC simulations, (c) without and (d) with return hops; $\psi^{\rm BT}(\tau)$ is converted from the static distribution $p^{\rm BT}_{\rm G-REM}(\Delta E)$ using the Arrhenius law.}
    \label{fig:energy_trap}
\end{figure}

\subsection*{2. Searching for landscape basins and barriers: barrier-tree (BT) algorithm}

The BT algorithm searches for all local minima and saddle points in the energy landscape, and organizes them into a barrier tree (see Fig.~\ref{fig:barrier}a for an example of a sub-tree). A spin configuration is referred to as a local minimum  if its energy is lower than the energy of any adjacent configuration (each configuration has $N$ adjacent configurations related by a single spin flip). The complete set of local minima are found by exhaustive search. To find the saddle point between two local minima, the algorithm first searches for all possible paths (a path is a series of subsequent spin flips) between the two minima, with  the maximum energy point identified along each path -- the saddle point is then defined as the lowest energy point among all maxima (min-max).

The barrier tree is constructed recursively in the following way: (i) find all $N_{\rm b}$ local minima; (ii) connect the two lowest local minima by a saddle point, and replace this sub-tree with a new node whose energy is equivalent to the saddle point energy (the new set has $N_{\rm b} -1$ nodes); (iii) repeat (i) and (ii) until only one node is left in the set ($N_{\rm b} = 1$).  More details about the  algorithm can be found in Refs.~\cite{flamm2002barrier, flamm2000rna, ferreira2000landscape, fontanari2002fractal}.

\begin{figure}
    \centering
    \includegraphics[width=1.\linewidth]{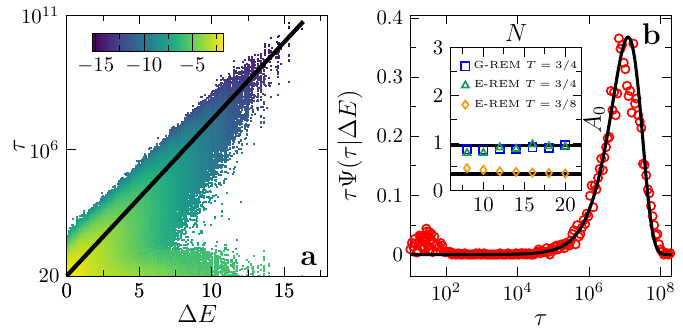}
    \caption{
{\bf Arrhenius law. } (a) Correspondence between $\tau$ and $\Delta E$ in MC simulations of the G-REM at {$T=0.75$} with $N=20$, where the color represents the logarithm of the joint probability $\ln \Psi(\tau, \Delta E)$ (see the color bar). (b) The conditional probability $\Psi(\tau | \Delta E=10)$ multiplied by $\tau$; the solid line follows Eq.~(\ref{eq:Psi}).   
(b-inset) $A_0$ vs $N$.
} 
    \label{fig:Arrhenius}
\end{figure}

As shown in Fig.~\ref{fig:barrier}a, generally a local minimum can be connected to multiple saddle points. For a given local minimum with an energy $E_{\rm lm}$,  we
define $E_{\rm sp}$ as the lowest energy of its connected saddle points, and the barrier energy is given by $\Delta E = E_{\rm sp} - E_{\rm lm}$.  
Our definition of $\Delta E$ is also consistent with the above min-max definition of  saddle points.
In this way, each local minimum is assigned to an energy barrier $\Delta E$. The probability distribution of $\Delta E$ gives $p^{\rm BT}(\Delta E)$.

Each glass basin corresponds to a local minimum and a saddle point defined in the above way. A glass basin also contains other configurations that are neither local minima nor saddle points --  they form a set of configurations connected to the local minimum whose energies are all below $E_{\rm sp}$. Numerically, we start from the given local minimum, and search for its direct neighbours with a single-spin flip, the neighbours of neighbours, ..., until the configuration's energy $E$ is larger than $E_{\rm sp}$. In this way, for a given realization of the REM  with a finite $N$, we find all $i =1, 2, \ldots, N_{\rm b}$ basins. Each basin corresponds to a set $\mathcal{B}_i$ of  configurations belonging to it, a local minimum energy $E_{\rm lm}^i$, a saddle point energy  $E_{\rm sp}^i$, and a barrier energy $\Delta E_i = E_{\rm sp}^i - E_{\rm lm}^i$. Note that many configurations do not belong to any basins.

\subsection*{3. Simulations of  single-spin flip Metropolis dynamics: Monte Carlo (MC) algorithm}

Dynamical trajectories, which are time sequences of configurations $\mathcal{C}(t)$,  are obtained by standard single-spin flip MC simulations, starting from random initial configurations. An example is provided in Fig.~\ref{fig:energy_trap}a, where the energy $E(t_{\rm w})$ of the configuration at $t_{\rm w}$ is plotted. In order to obtain the aging function $\Pi_w(t_{\rm w})$, our task is to transform the configuration trajectory to a basin trajectory (Fig.~\ref{fig:energy_trap}b). The detailed procedure is described below.

During the dynamics, the system is in the basin $\mathcal{B}_i$ if $\mathcal{C}(t) \in \mathcal{B}_i $ and it leaves the basin if $\mathcal{C}(t) \notin \mathcal{B}_i $ anymore. The duration in the basin defines the  hopping time (trapping time) $\tau$. In this way, we identify a sequence of hopping events with $\{\tau_k, E_{\rm lm}^{i_k}, E_{\rm sp}^{i_k}, \Delta E_{i_k}\}$, where $k=1,2,\cdots$ and $1\leq i_k \leq N_{\rm b}$ denotes that the $k$-th event is in  the basin $i_k$ (Fig.~\ref{fig:energy_trap}b). Note that the  transient time between basins is negligible, i.e., we assume that once the system leaves a basin, it immediately falls into the next basin.

A major computational challenge is the exponential growth of the size $\mathcal{O}(2^N)$ of the configuration space with the system size $N$.
For systems with $N>20$ spins, storing the entire landscape would exceed the memory limit of standard computers. To address this issue, we only save the portion of the landscape that is actually explored during the MC dynamical simulations.  This trick allows us to simulate up to $N=128$ spins for maximally
$10^9$ MC steps, with 1~GB memory.

\subsection*{4. Test of the renewal mechanism}

Occasionally, after leaving a basin, the system jumps back to the same basin. 
Following the previous study~\cite{baity2018activated}, the two events are merged in such cases. The effects of return jumps are revealed in {Fig.~\ref{fig:energy_trap}(c,d)} by the conditional probability  $p(\tau_2 | \tau_1)$, which is the distribution of the next-event hopping time $\tau_2$ when the current-event hopping time is $\tau_1$. Only when the return jumps are merged, the $p(\tau_2 | \tau_1)$ is independent of $\tau_1$, which means that the two consecutive jumps are independent. The independence of consecutive jumps (the renewal mechanism) is a basic assumption in the trap model. In order to be 
 consistent with the renewal mechanism, we employ the no-return treatment.

\subsection*{5. Verification of the Arrhenius law} 

In the trap model, the hopping time $\tau$ and the barrier energy $\Delta E$ are related through the
 Arrhenius law, 
\beq
\tau(\Delta E) = A_0 \exp(\Delta E/T).
\label{eq:Arrhenius}
\eeq 
We find that the Arrhenius law is consistent with  the  $\bar{\tau}(\Delta E)$ data ($\bar{\tau}$ is the mean hopping time for the given $\Delta E$) obtained by MC simulations (Fig.~\ref{fig:Arrhenius}a), with $A_0 \sim O(1)$ as shown in Fig.~\ref{fig:Arrhenius}b-inset (the time unit is $N$).

When $N$ is finite, for a given $\Delta E$, $\tau$ follows a distribution 
\beq
\Psi(\tau | \Delta E) = (1- \eta)^{\tau-1} \eta \approx e^{-\tau \eta} \eta
\label{eq:Psi},
\eeq
where $\eta = \exp(-\Delta E/T)$.
To derive $\Psi(\tau | \Delta E)$, consider a discretized time $\tau$: if a system is trapped in a basin with a  $\Delta E$-barrier for $\tau$ steps, it should remain in the basin for $\tau - 1$ steps with a probability $1-\eta$ at each step, and jumps out the basin at the final step with a probability $\eta$. Equation~(\ref{eq:Psi}) is also verified by the simulation data (see Fig.~\ref{fig:Arrhenius}b).
Note that the mean hopping time  of Eq.~(\ref{eq:Psi}), $\bar{\tau}(\Delta E) = \int_0^\infty \tau \Psi(\tau | \Delta E) d\tau$, consistently recovers the Arrhenius law Eq.~(\ref{eq:Arrhenius}). In short, the Arrhenius law is verified by our MC data. 

\section*{Appendix B: weak ergodicity breaking order parameter}

\begin{figure}
    \centering
    \includegraphics[width=\linewidth]{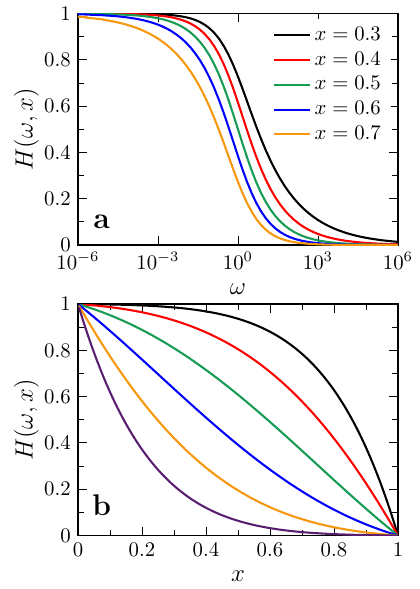}
    \caption{{\bf WEB order parameter $H(w,x)$ by the arcsin law Eq.~(\ref{eq:Pi}).} 
    (a) $H$ versus $w$ for a few different $x$.
    (b) $H$ versus $x$ for $\omega=0.01,0.1,0.5,2,10,100$ (top to bottom). 
    }
    \label{fig:hwx}
\end{figure}

Figure \ref{fig:hwx} shows how $H(w,x)$ depends on $w$ for several given $x$, and on $x$ for several given $w$, as generalization of the theoretical curves in Fig.~\ref{fig:aging}d (blue line) and Fig.~\ref{fig:theory}a.

\section*{Appendix C: Weeks-Chandler-Andersen (WCA) model}\label{appendix:wca}

\subsection*{1. Model}

The WCA model is an $80:20$ binary mixture of type A and type B particles interacting with the Lennard-Jones (LJ) potential, 
\beq
V(r) = 4\varepsilon_{\alpha\beta}\left[(\frac{r}{\sigma_{\alpha\beta}})^{12} - (\frac{r}{\sigma_{\alpha\beta}})^6 \right],
\eeq
where $\alpha,~\beta\in\{A,~B\}$, with parameters $\sigma_{\rm AB}/\sigma_{\rm AA} = 0.8$, $\sigma_{\rm BB}/\sigma_{\rm AA} = 0.88$, $\varepsilon_{\rm AB}/\varepsilon_{\rm AA} = 1.5$, and $\varepsilon_{\rm BB}/\varepsilon_{\rm AA} = 0.5$. 
The pair potential $V(r)$ is truncated and shifted to zero at the minimum, $r^{\rm cut}_{\alpha\beta} = 2^{1/6}\sigma_{\alpha\beta}$. 
All particles have the same unit mass, $m_{\rm A} = m_{\rm B} = m$. The parameters $\varepsilon_{\rm AA}$, $\sigma_{\rm AA}$ and $\sqrt{m\sigma_{\rm AA}^2/\varepsilon_{\rm AA}}$ are used as the energy, length, and time units. {The presented data are obtained  for  large (type $\rm A$) particles.}

\subsection*{2. Molecular dynamics  (MD) simulation method}

Simulations of the WCA model are performed at a fixed number density $\rho = 1.2$. The MCT temperature at this density is $T_{\rm MCT}  \approx 0.28$~\cite{role2011berthier}. 
We generate equilibrium configurations at an initial temperature $T_{\rm eq} = 5.0$, and then instantly quench the samples to a target temperature $T$.
To suppress the wild temperature fluctuations at the initial stage, the aging simulations are performed under the isokinetic ensemble, a variant of the canonical ensemble~\cite{tuckerma2015statisti}.
The temperature (kinetic energy) is rescaled to the target temperature every 10000 MD steps (100 LJ time units).

\begin{figure}[!htbp]
    \includegraphics[width=\linewidth]
    {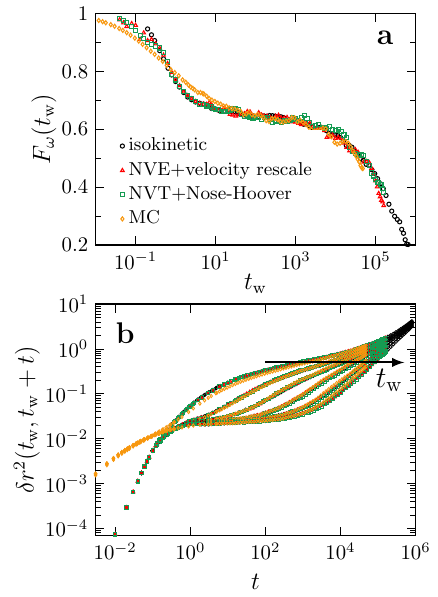}
    \caption{{\bf Comparison of the results obtained by different MD and MC algorithms for the WCA model.}  
(a) Aging function $F_w(t_{\rm w})$ and 
(b) non-equilibrium two-time  MSD  $\delta r^2(t_{\rm w}, t_{\rm w}+t)$ for $N=1000$ particles at $T=0.25$.
In panel (b),   $t_{\rm w} \approx 1.0~,16.0,~135,~1.15\times10^3,~9.65\times10^{3},~4.21\times10^{4}$ from left to right.
}
\label{fig:compare_WCA}
\end{figure}

We have checked that the aging functions $F_w(t_{\rm w})$  obtained by different simulation methods are consistent with each other, apart from the short-time dynamics that are protocol-dependent (see Fig.~\ref{fig:compare_WCA}a).
{Four different simulation methods are compared. (i) MD simulations in the isokinetic ensemble. (ii) MD simulations in the microcanonical (NVE) ensemble with the kinetic energy rescaled to the target value every 50 MD steps (0.1 LJ time units). (iii) MD simulations in the canonical (NVT) ensemble using the Nose-Hoover chain thermostat with  a damping parameter of 0.1 LJ time units~\cite{tuckerma2015statisti}. (iv) Standard MC simulations in the canonical ensemble. In the MC simulation, we attempt to translate a particle by a displacement vector randomly drawn in a sphere of a radius $5\times10^{-2}\sigma_{\rm AA}$. The move is accepted according to the Metropolis acceptance rule. 
In Fig.~\ref{fig:compare_WCA}, the MC steps are multiplied  by $6.0\times10^{-4}$ to match the long-time dynamics with those obtained by MD simulations.}
Further comparison is carried out for the non-equilibrium two-time  mean-squared displacement (MSD), $\delta r^2(t_{\rm w}, t_{\rm w}+t) \equiv \frac{1}{N_{\rm A}} \langle \sum_{i=1}^{N_{\rm A}} |{\bf r}_i(t_{\rm w}+t) - {\bf r}_i(t_{\rm w})   |^2\rangle$ (see Fig.~\ref{fig:compare_WCA}b).
These tests  validate the isokinetic ensemble adapted in our MD simulations, which is used to generate  the results reported in the current study.

\section*{Appendix D: Amorphous silica model}
\label{appendix:sio2}

\subsection*{1. Model}

The atomic interactions in the amorphous silica ($\rm{SiO}_2$) is modeled by  the Beest Kramer van Santen (BKS) potential, derived from ab initio calculations and lattice dynamics calculations~\cite{van1990force}. The  BKS potential for both $\rm{Si-O}$ and $\rm{O-O}$ interactions diverges to negative infinity at a small distance $r$, leading to unrealistic attractions  at high temperatures.  To address this issue, the BKS potential at small distances is replaced by a harmonic repulsive potential, which gives the interactions in the following form:
\[
V(r) =
\begin{cases}
D_{\alpha\beta}(r - r_{\alpha\beta}^c)^2, & r < r_{\alpha\beta}^c, \\
\displaystyle\frac{q_{\alpha} q_{\beta} e^{2}}{r} + A_{\alpha\beta} \exp(-B_{\alpha\beta} r) - \frac{C_{\alpha\beta}}{r^6}, & r \geq r_{\alpha\beta}^c,
\end{cases}
\]
where $\alpha, \beta \in \{\mathrm{Si}, \mathrm{O}\}$.
The parameters of BKS potential can be found in the standard reference~\cite{van1990force}.
For the harmonic potential, $D_{\alpha\beta} = 100$,   $r_{\rm{OO}}^c=1.43869$ and $r_{\rm{SiO}}^c=1.19362$.  The Coulomb interaction is computed using the Ewald method. All quantities are expressed in metal units.

\subsection*{2. Molecular dynamics  simulation method}

The MD simulations are carried out using the Large-scale Atomic/Molecular Massively Parallel Simulator (LAMMPS)~\cite{thompson2022lammps},  at a constant volume with a fixed mass density $\rho_{\rm m}=2.36 \rm{g/{cm}^3}$. The Verlet algorithm with a time step of 1.6 $\rm{fs}$ is used to integrate the equations of motion. We simulate the aging dynamics by quenching the system from a high initial temperature $T_{\rm eq}=8000~{\rm K}$ to a low target temperature $T$. The initial equilibrium configurations at $T_{\rm eq}$ are melted from a crystalline $\rm{SiO}_2$ in the NVT ensemble. After quenching, the aging dynamics are simulated  with velocities rescaled  every 50 steps to keep the target temperature $T$.

\subsection*{3. Estimate of the onset temperature $T_{\rm onset}$}

\begin{figure}[!htbp]
    \includegraphics[width= \linewidth]{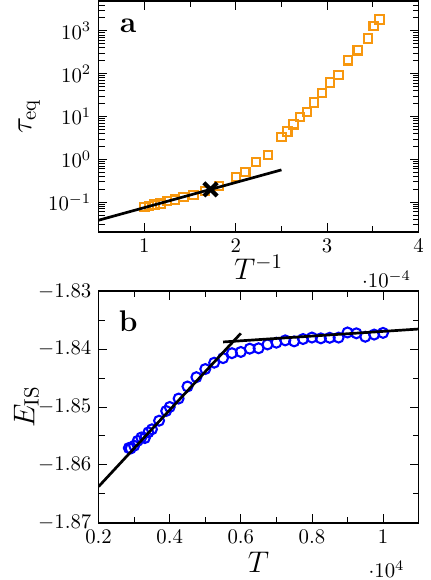}
\caption{{\bf Onset temperature $T_{\rm onset}$ in the amorphous silica model.}
(a) $\alpha$-relaxation time $\tau_{\rm eq}$  as a function of $1/T$. The high-$T$ data are fitted  to the Arrhenius law (line), and the departing point indicates $T_{\rm onset}$.
(b) Inherent structure energy $E_{\rm IS}$ as a function of $T$. The two lines represent linear fitting to low-$T$ and high-$T$ data, whose intersection defines $T_{\rm onset}$. Data are obtained for $N=1536$.
}
\label{fig:T_onset_silica}
\end{figure}

The onset temperature $T_{\rm onset}$ is defined as the characteristic temperature below which glassy dynamics appear.  
For amorphous silica,  $T_{\rm onset}$ has not yet been reported  in the literature. In this study two independent methods are employed to estimate  $T_{\rm onset}$. The first method is based on the $T$-dependent behavior of the $\alpha$-relaxation time $\tau_{\rm eq}(T)$ in supercooled liquids.  One defines  
$T_{\rm onset}$ at the point where $\tau_{\rm eq}(T)$ departs from the high-$T$ Arrhenius behavior, $\tau_{\rm eq}(T) \sim \exp(E_\infty/T)$. Applying this criterion to our simulation data gives {$T_{\rm onset} \approx 5800$}~K (see Fig.~\ref{fig:T_onset_silica}a). Note that for amorphous silica, $\tau_{\rm eq}(T)$ is also Arrhenius  at low $T$ (with a constant, but larger activation energy than $E_\infty$), consistent with the well-know fact that amorphous silica forms strong glasses.

The second method is based on the $T$-dependent behavior of the potential energy $E_{\rm IS}(T)$ of inherent structures~\cite{brumer2004mean}.  
The inherent structures at zero temperature are obtained by 
minimizing the energy of  configurations initially equilibrated at $T$ using the FIRE algorithm~\cite{bitzek2006structural}. The high-$T$ and low-$T$ data of $E_{\rm IS}(T)$ are fitted by two linear functions, whose intersection defines {$T_{\rm onset} \approx 5800$}~K (see Fig.~\ref{fig:T_onset_silica}b). The above two methods give a consistent  $T_{\rm onset}$ within the numerical uncertainty.

\section*{Appendix E: nonergodicity order parameter in structural glasses}

\begin{figure}[!htbp]
    \centering
    \includegraphics[width=\linewidth]{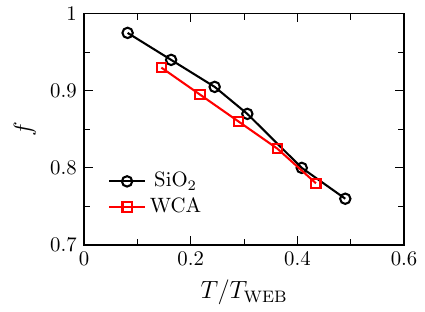}
    \caption{{\bf Nonergodicity parameter $f$ with respect to the rescaled temperature $T/T_{\rm WEB}$, for WCA and amorphous silica models.}}
    \label{fig:f-T}
\end{figure}

The nonergodicity  parameter $f$ is defined by $f = \lim_{t\to \infty} F_{\rm eq}(t)$, where $F_{\rm eq}(t)$ is the equilibrium incoherent
scattering function. For the aging data,  $f$ corresponds to the plateau value of $F(t_{\rm w}, t_{\rm w}+t)$ by sending $t_{\rm w} \to \infty$ first (equilibrium limit) and then $t \to \infty$ (large-time limit); see Fig.~\ref{fig:aging}c. 
 Figure~\ref{fig:f-T} shows the data of $f$ obtained for both WCA and amorphous silica models. Interestingly, the data nearly collapse if plotted as functions of the rescaled temperature $T/T_{\rm WEB}$. 

\section*{Appendix F: Exponent $\beta$ in structural glasses}

\begin{figure}[!htbp]
    \includegraphics[width= \linewidth]{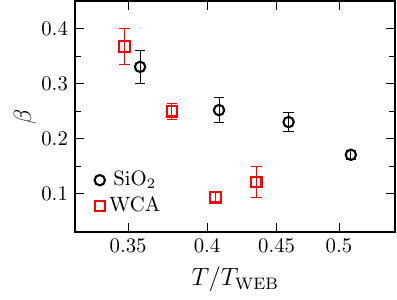}
\caption{{\bf Exponent $\beta$ as a function of $\hat{x} = T/T_{\rm WEB}$ in WCA and amorphous silica models.
} 
}
\label{fig:beta}
\end{figure}

According to our theory, the aging function  follows logarithmic behavior in the intermediate-time regime, $F_w(t_{\rm w}) \sim -k \ln t_{\rm w}$. For mean-field models (e.g., REMs), $k\sim 1/N$ (see Eq.~\ref{eq:scaling_GTM}). For finite-dimensional models, such as the WCA and amorphous silica models, we find that $k \sim N^{-\beta}$ as shown in Fig.~\ref{fig:WCA}(c, f) (in the WCA model, $k$ becomes a constant when $N>n$). The $T$-dependent exponent $\beta$ is reported in Fig.~\ref{fig:beta}. For both models, $\beta \approx 0.1-0.4$, which is considerably smaller than the mean-field value $\beta = 1$. It is unclear if this discrepancy comes from finite-dimensional effects or pre-asymptotic effects due to insufficiently large systems.

\bibliography{aging}
\clearpage

\onecolumngrid

\centerline{\Large \bf Supplemental Material}

\setcounter{figure}{0}
\setcounter{equation}{0}
\setcounter{table}{0}
\setcounter{section}{0}
\renewcommand\thefigure{S\arabic{figure}}
\renewcommand\theequation{S\arabic{equation}}
\renewcommand\thesection{S\arabic{section}}
\renewcommand\thetable{S\arabic{table}}

\section{Aging theory of the generalized trap model}

\subsection{Outline of the general strategy}

Our theoretical derivation of the behavior of the aging function $\Pi(t_{\rm w}, t_{\rm w} + t)$ follows the strategy provided in Refs.~\cite{bouchaud1995aging, monthus1996models}. If the system is trapped in a basin with a trapping time $\tau$, then the probability of the system remaining in the basin after a time $t$ decays exponentially as $\sim e^{-t/\tau}$. According to this, one can write,
\beq
\Pi(t_{\rm w}, t_{\rm w}+t)  = \left \langle 
\sum_{\beta=1}^{N_{\rm b}}
 Q(t_{\rm w}, \tau_\beta )  e^{-t/\tau_\beta} \right \rangle =
N_{\rm b} \int_{\tau_0}^\infty d\tau {Q}(t_{\rm w}, \tau) e^{-t/\tau} \psi(\tau),
\label{eq:Laplace_PI}
\eeq
where $N_{\rm b}$ is the total number of basins,  $Q(t_{\rm w}, \tau)$ is the probability of the system being in a basin with a hopping time $\tau$ at a waiting time $t_{\rm w}$,  $\psi(\tau)$ is the hopping time probability distribution function, and $\tau_0$ is the minimum hopping time (in the main text, we have set $\tau_0=1$ as the unit of time). The  distribution $\psi(\tau)$ is normalized as,
\beq
\int_{\tau_0}^\infty  d\tau \psi(\tau) = 1.
\label{eq:normalization}
\eeq
In the original Bouchaud’s trap model (BTM), the energy barrier distribution is exponential (see Eq.~1). Using the Arrhenius law $\tau(\Delta E) \sim \exp(\Delta E/T)$, the exponential barrier distribution  $P(\Delta E) \sim \exp(-\Delta E/T_{\rm c})$ can be converted to a power-law hopping time distribution  $\psi(\tau)$,
\beq
\psi(\tau) = x \tau_0^x \tau^{-(1+x)},
\label{eq:psi_tau_1}
\eeq
where  $x = T/T_{\rm c}$. Note that Eq.~(\ref{eq:psi_tau_1})
 is normalized according to Eq.~(\ref{eq:normalization}). 

Bouchaud et al. showed that it is convenient to work in the Laplace transform of $Q(t_{\rm w}, \tau_\beta )$, 
\beq
\hat{Q}(s, \tau) = \mathcal L\{s Q(t_{\rm w}, \tau)\} 
= \int_{0}^\infty dt_{\rm w} e^{-s t_{\rm w} } s Q(t_{\rm w}, \tau).
\label{eq:hatQ1}
\eeq
When the consecutive hops are independent, the probability $\hat{Q}(s, \tau)$ is~\cite{bouchaud1995aging, monthus1996models},
\beq
\hat{Q}(s, \tau) = \frac{\frac{s\tau}{s\tau+1}}{\mathcal{C} N_{\rm b} \left \langle \frac{s\tau}{s\tau+1} \right \rangle},
\label{eq:hatQ2}
\eeq
where $\mathcal{C}$ is a normalization constant. If one considers $t_{\rm w}$ as an exponentially distributed random variable, as suggested by Eq.~(\ref{eq:hatQ1}), then Eq.~(\ref{eq:hatQ2}) can be understood as $\hat{Q}(s, \tau) \sim \tau/(\tau + \overline{t}_{\rm w})$,  where the mean waiting time $\overline{t}_{\rm w} \sim 1/s$. It suggests that the probability of being in a basin with a hopping time $\tau$ increases with $\tau$ and decreases with the  waiting time $\overline{t}_{\rm w}$.
Let us consider several limiting cases. 
For any finite $\tau$, $\hat{Q}(s, \tau) \to 1$ when $\overline{t}_{\rm w} \to 0$ (the probability of escaping  from any basin is zero without waiting), and $\hat{Q}(s, \tau) \to 0$ when $\overline{t}_{\rm w} \to \infty$ (all basins can be escaped from after long-time waiting). For any finite $\overline{t}_{\rm w}$, $\hat{Q}(s, \tau) \to 0$ when $\tau \to 0$ (the probability to stay in the basin with $\tau=0$ is zero), and  $\hat{Q}(s, \tau) \to 1$ when $\tau \to \infty$ (the basin can not be escaped from if its $\tau$ is infinite). 

With Eqs.~(\ref{eq:Laplace_PI}),~(\ref{eq:psi_tau_1}),~(\ref{eq:hatQ1}) and~(\ref{eq:hatQ2}), the aging function $\Pi(t_{\rm w}, t_{\rm w}+t)$ can be obtained. For the generalized trap model (GTM), Eq.~(\ref{eq:psi_tau_1}) needs to be modified in order to include the finite-size correction, but the other equations can be kept. In the following analyses, the results in Sections.~\ref{sec:scaling} and~\ref{sec:short-time} are general for long-time and short-time dynamics in the BTM and GTM, while the  logarithmic decay discussed in Section~\ref{sec:log} is uniquely due to the finite-size effects in the GTM.

\subsection{Long-time power-law convergence to the asymptotic plateau}
\label{sec:scaling}
Here we give a theoretical derivation of the second line ($\tau_{\rm m} < t_{\rm w} < \tau_{\rm th}$) in
Eq.~(5) of the main text. 
In Eq.~(\ref{eq:hatQ2}), the normalization constant $\mathcal{C}=1$ in the large-time limit $t_{\rm w} \to \infty$. With $\mathcal{C}=1$, the plateau $H(w,x)$ of the aging function can be determined as given by the arcsin law Eq.~(2). In order to obtain the behavior of how the aging function converges to the plateau, it is crucial to consider the next-order $1/t_{\rm w}$-correction to $\mathcal{C}$, as shown below. Note that in this section, we use the BTM distribution Eq.~(\ref{eq:psi_tau_1}) without finite-size corrections. With Eq.~(\ref{eq:psi_tau_1}), the average in the denominator of Eq.
(\ref{eq:hatQ2}) becomes,
\beq
 \left \langle \frac{s\tau}{s\tau+1} \right \rangle = \int_{\tau_0}^\infty d\tau \frac{s\tau}{s\tau+1}  \psi(\tau) 
 = \prescript{ }{2}{F}_1(1,x;1+x;-\frac{1}{s\tau_0})
 , 
 \label{eq:average}
\eeq
where $\prescript{ }{2}{F}_1(a,b;c;z)$ is the ordinary hypergeometric function. In the large waiting time limit $t_{\rm w} \to \infty$ (or equivalently in the limit $s \to 0$), 
 {$\prescript{ }{2}{F}_1(1,x;1+x;-\frac{1}{s\tau_0})
 \approx x \tau_0^{x}s^{x} \frac{\pi}{\sin(\pi x)}$.}
 Thus an approximate expression of $\hat{Q}(s, \tau)$  is,
 \beq
\hat{Q}(s, \tau)  \approx  \frac{\sin( \pi x) \tau_0^{-x}}{\mathcal{C} N_{\rm b} \pi x} \frac{s\tau}{(s\tau+1)}s^{-x}.
\label{eq:hatQ_1}
 \eeq
The probability distribution $Q(t_{\rm w}, \tau)$ can be obtained by performing an inverse Laplace transform to Eq.~(\ref{eq:hatQ_1}) and then applying the convolution theorem,
\beq\label{eq:Qtwtau}
{Q}(t_{\rm w}, \tau) &= \mathcal{L}^{-1}\left\{\frac{1}{s}\hat{Q}(s, \tau) \right \}
\approx \frac{\sin( \pi x) \tau_0^{-x}}{\mathcal{C} N_{\rm b} \pi x \Gamma(x)} \int_0^{t_{\rm w}} dt' (t_{\rm w}-t')^{x-1} e^{-t'/\tau},
\eeq
where $\Gamma(x)$ is the gamma function.
 The normalization condition requires that,
 \beq
\left \langle \sum_{\beta}^{N_{\rm b}}
 {Q}(t_{\rm w}, \tau_\beta ) \right \rangle = N_{\rm b} \int_{\tau_0}^\infty d\tau {Q}(t_{\rm w}, \tau ) \psi(\tau) = 1,
 \eeq
or,
\beq
\mathcal{C} = \frac{\sin( \pi x) }{\pi \Gamma(x)}   
& \times \int_{\tau_0}^\infty d \tau \int_0^{t_{\rm w}} dt' \tau^{-(1+x)} (t_{\rm w}-t')^{x-1} e^{-t'/\tau}.
\label{eq:C1}
\eeq
Note that $\mathcal{C}$ depends on $t_{\rm w}$. Plugging Eqs.~(\ref{eq:psi_tau_1}) and~(\ref{eq:Qtwtau}) into Eq.~(\ref{eq:Laplace_PI}), we obtain, 
\beq
\begin{split}
\Pi(t_{\rm w}, t_{\rm w}+t) &\approx \frac{\sin( \pi x) }{\mathcal{C} \pi \Gamma(x)}   \int_0^\infty d \tau \int_0^{t_{\rm w}} dt' \tau^{-(1+x)} (t_{\rm w}-t')^{x-1} e^{-(t'+t)/\tau}\\
&\approx \frac{\sin( \pi x) }{\mathcal{C} \pi}    \int_0^{t_{\rm w}} dt' 
\left( \frac{t+t'}{t_{\rm w} -t'} \right)^{-x} (t_{\rm w}-t')^{-1}
\\
&= \frac{\sin( \pi x) }{\mathcal{C} \pi}    \int_{w}^{\infty} du 
u^{-x} (1+u)^{-1}, 
\end{split}
\label{eq:Pi1}
\eeq
where  $u = \frac{t+t'}{t_{\rm w} - t'}$ and $w = t/t_{\rm w}$. 

\begin{figure}[!htbp]
  \centering
  \includegraphics[scale=0.95]{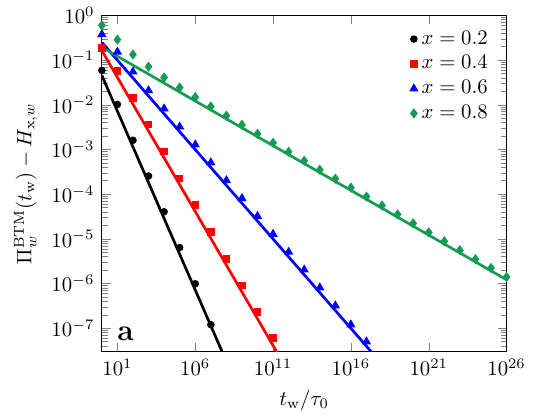}
  \includegraphics[scale=0.95]{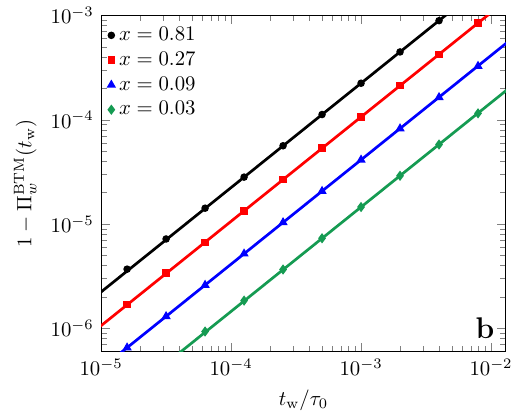}
  \caption{
{\bf Comparison between the analytical approximation of the aging function for the BTM and the exact numerical results using the inverse Laplace transform.} 
(a) Large-$t_{\rm w}$ power-law convergence to the asymptotic plateau. (b) Small-$t_{\rm w}$ linear decay. 
The points are exact results obtained by the numerical inverse Laplace transform of Eqs.~(\ref{eq:hatQ2}) and ~(\ref{eq:average}). The solid lines are Eq.~(\ref{eq:Pi2}) in (a) and Eq.~(\ref{eq:lineardecay}) in (b).
}
  \label{fig:power_law_decay}
\end{figure}

To the zeroth order of $t_{\rm w}$, Eq.~(\ref{eq:C1}) gives $\mathcal{C} \approx 1$, and then  Eq.~(\ref{eq:Pi1}) recovers the arcsin law 
Eq.~(2). Expanding  Eq.~(\ref{eq:C1}) to the next order of $1/t_{\rm w}$, we obtain,
\beq
\begin{split}\label{eq:hatPi1}
\mathcal{C} &= 1 - \frac{\sin(\pi x)}{\pi \Gamma(x)}\int_{0}^{\infty} du (1+u)^{-1} u^{-x} \Gamma\left(x,\frac{u}{1+u} \frac{t_{\rm w}}{\tau_0} \right) \\
&\approx1 - \frac{1}{(1-x)\Gamma(1-x)\Gamma^2(x)}\left(\frac{t_{\rm w}}{\tau_0} \right)^{x-1}.
\end{split}
\label{eq:C2}
\eeq
Combing Eqs.~(\ref{eq:Pi1}),~(\ref{eq:C2}) and~(2) gives, 
\beq
\begin{split}
\Pi_w^{\rm BTM}(t_{\rm w})
& \approx  H(w,x) \left[1 + \frac{1}{(1-x)\Gamma(1-x)\Gamma^2(x)}\left(\frac{t_{\rm w}}{\tau_0} \right)^{x-1}\right]\\
& = H(w,x) \left[1 + A \left(\frac{t_{\rm w}}{\tau_0} \right)^{-\alpha}\right],
\end{split}
\label{eq:Pi2}
\eeq
where $\alpha = 1-x$ and  $A =  \frac{1}{(1-x)\Gamma(1-x)\Gamma^2(x)}$.
We have thus derived the large-$t_{\rm w}$ form of 
Eq.~(5). Finally, we compare the analytic expression Eq.~(\ref{eq:Pi2}) with the numerical results obtained by the inverse Laplace transform of the exact expressions Eqs.~(\ref{eq:hatQ2}) and ~(\ref{eq:average}), confirming the validity of  Eq.~(\ref{eq:Pi2}) in the asymptotic regime (see Fig.~\ref{fig:power_law_decay}a).  Note that, without loss of generality, we fix $w=0.5$ in this study. 

\subsection{Short-time linear decay}
\label{sec:short-time}
For short-time dynamics, similarly we begin with Eq.~(\ref{eq:average}) in the last section.
In the short-time limit $t_{\rm w}\rightarrow 0$ (corresponding to the limit $s\rightarrow\infty$),
$\prescript{ }{2}{F}_1(1,x;1+x;-\frac{1}{s\tau_0})
 \approx 1$,
 and
\beq
\hat{Q}(s,\tau)\approx \frac{1}{\mathcal{C}N_{\rm b}}\frac{s\tau}{s\tau+1}.
\eeq
The probability distribution $Q(t_{\rm w},\tau)$ is again obtained by the inverse Laplace transform:
\beq
Q(t_{\rm w},\tau) = \frac{1}{\mathcal{C}N_{\rm b}}e^{-t_{\rm w}/\tau}.
\eeq
Then the aging function is obtained, 
\beq
\begin{split}\label{eq:pi2}
\Pi(t_{\rm w},t_{\rm w}+t)
&= N_{\rm b}\int_{\tau_0}^\infty d\tau Q(t_{\rm w},\tau)e^{-t/\tau}\psi(\tau)\\
&= \frac{x}{\mathcal{C}}\left( \frac{t+t_{\rm w}}{\tau_0 }\right)^{-x} \left[\Gamma(x) - \Gamma\left(x,\frac{t+t_{\rm w}}{\tau_0}\right)\right]\\
&= \left( \frac{t_{\rm w}}{t+t_{\rm w}} \right)^x\frac{\Gamma(x) - \Gamma\left(x,\frac{t+t_{\rm w}}{\tau_0}\right)}{ \Gamma(x) - \Gamma\left(x,\frac{t_{\rm w}}{\tau_0}\right) },\\
\end{split}
\eeq
where the normalized constant is,
\beq\label{eq:normal2}
\begin{split}
\mathcal{C} &= \int_{\tau_0}^\infty d\tau e^{-t_{\rm w}/\tau} \psi(\tau)\\
&= x\left(\frac{t_{\rm w}}{\tau_0}\right)^{-x}  \left[ \Gamma(x) - \Gamma\left(x,\frac{t_{\rm w}}{\tau_0}\right) \right ].
\end{split}
\eeq
Expanding Eq.~(\ref{eq:pi2}) around $t_{\rm w}=0$, we obtain,
\beq\label{eq:lineardecay}
\Pi_w^{\rm s}(t_{\rm w}) \approx 1 - \frac{w x}{x+1}\left( \frac{t_{\rm w}}{\tau_0} \right) =  1 - C \left( \frac{t_{\rm w}}{\tau_0} \right).
\eeq
Similar to the long-time case, the analytic result Eq.~({\ref{eq:lineardecay}) can be confirmed by exact numerical calculations  using the inverse Laplace transform (see Fig.~\ref{fig:power_law_decay}b).

\subsection{Logarithmic decay due to finite-size effects in the generalized trap model}
\label{sec:log}

The GTM aging function is computed using the formula Eq.~(\ref{eq:Laplace_PI}),
\beq
\begin{split}
\Pi_w^{\rm GTM}(t_{{\rm w}}) & =\int_{\tau_0}^{\infty}d\tau\psi_{\rm GTM}(\tau)Q_{\rm GTM}(t_{{\rm w}},\tau)e^{-w t_{{\rm w}}/\tau}\\
 & =\int_{0}^{\infty}d\tau e^{f_{\rm GTM}(t_{{\rm w}},\ln \tau)}.
 \end{split}
 \label{eq:PI_GTM}
\eeq
The GTM barrier energy distribution Eq.~(3) leads to 
 the modified trapping time distribution (via the Arrhenius law), 
\beq
\begin{split}
\psi_{\rm GTM}(\tau)
&\sim \tau^{-\hat{x}-1} \exp\left[- \mu \ln^2(\tau/\tau_0) \right] \\
&\sim \tau^{-\hat{x}-\mu\ln(\tau/\tau_{0})-1}, \\
\end{split}
\eeq
where $\mu\equiv \frac{T^2}{2Na}$,
and  $\hat{x}= \left(1-\frac{\bar{E} T_{\rm c}}{Na} \right)x$.
The $Q_{\rm GTM}(t_{\rm w}, \tau)$ can be computed by the inverse Laplace transform of Eq.~(\ref{eq:hatQ2}), where $
 \left \langle \frac{s\tau}{s\tau+1} \right \rangle = \int_{\tau_0}^\infty d\tau \frac{s\tau}{s\tau+1}  \psi_{\rm GTM}(\tau)$.

 It is hard to directly analyze Eq.~(\ref{eq:PI_GTM}). However, based on a saddle-point approximation, we find that the aging functions in the two models can be related by shifting the effective reduced temperature,
 \beq
\Pi_w^{{\rm GTM}}(t_{{\rm w}};x) \sim\Pi_w^{{\rm BTM}}(t_{{\rm w}};\tilde{x}),
\label{eq:Pirelation}
\eeq
where 
\beq 
\tilde{x} = \left(1-\frac{\bar{E} T_{\rm c}}{Na} \right)x +\frac{T^2}{Na}\ln t_{{\rm w}} = \hat{x} + 2\mu\ln t_{{\rm w}}.
\label{eq:tidlexeff}
\eeq

In other words, the aging function $\Pi_w^{{\rm GTM}}(t_{{\rm w}};x)$ of the GTM at a temperature $T = T_{\rm c}x$ is approximately equivalent to the aging function $\Pi_w^{{\rm BTM}}(t_{{\rm w}};\tilde{x})$ of the BTM at the temperature $T=T_{\rm c} \tilde{x}$ with $\tilde{x}$ given by Eq.~(\ref{eq:tidlexeff}).
Equation~(\ref{eq:Pirelation}) is essentially due to the relation between the maximum of $f_{\rm GTM}(t_{\rm w}, \ln \tau)$ in Eq.~(\ref{eq:PI_GTM}) and that of $f_{\rm BTM}(t_{\rm w}, \ln \tau)$,
\beq
\ln \tau^*_{\rm GTM}(t_{\rm w}, x) \approx \ln \tau^*_{\rm BTM}(t_{\rm w}, \tilde{x}),
\label{eq:tau}
\eeq
where $f(t_{\rm w}, \ln \tau)$ is maximized at $\ln \tau^*$ with other parameters ($t_{\rm w}$, $T$, etc.) fixed. Equation~(\ref{eq:tau}) is consistent with the  numerically evaluated maximum point $\ln \tau^*$ (see Fig.~\ref{fig:tau}). As  shown by the data, in general $\ln \tau^* \sim \ln t_{\rm w}$ in both models (Fig.~\ref{fig:tau}a,b).  
The GTM data at $x$ and the BTM data at the corresponding $\tilde{x}$ (see Eq.~\ref{eq:tidlexeff}) collapse for different $x$ and $N$  (Fig.~\ref{fig:tau}c,d), and thus Eq.~(\ref{eq:tau}) is verified.

\begin{figure}
    \centering
    \includegraphics[width=0.8\linewidth]{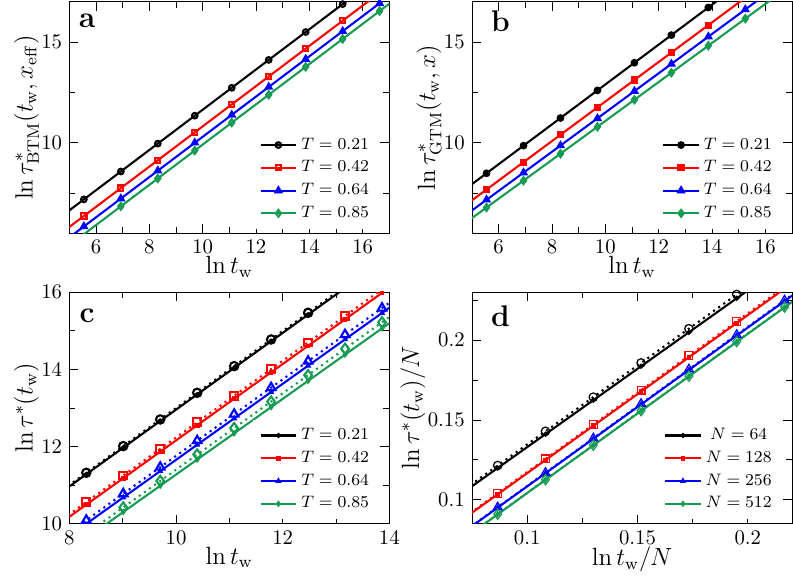}
    \caption{{\bf Verification of Eq.~(\ref{eq:tau}).} The data points are obtained by numerical evaluation of the maximum $f(t_{\rm w}, \ln \tau^*)$ of the function $f(t_{\rm w}, \ln \tau)$ defined in Eq.~(\ref{eq:PI_GTM}).
    As shown by the data, $\ln \tau^*$ is linearly proportional to $\ln t_{\rm w}$ in (a) BTM and (b) GTM, for $T=0.21, 0.42, 0.64, 0.85$ (or $x=0.25, 0.5, 0.75, 1.0$), and $N=64$.  At the same $T$, the BTM and GTM data do not collapse in (a,b).  However, the GTM data (dotted point-line) at $T=T_{\rm c} x$ and the BTM data (solid point-line) at $T = T_{\rm c} \tilde{x}$ collapse, where $\tilde{x}$ is given by Eq.~(\ref{eq:tidlexeff}), for different (c) $T$ (with $N=128$ fixed) and (d) $N$ (with $T=0.42$ fixed).
}
    \label{fig:tau}
\end{figure}

Equation~(\ref{eq:tidlexeff}) means that, the Gaussian term in the GTM barrier distribution Eq.~(3) gives rise to two effects compared to the BTM. First, it shifts the reduced temperature effectively from $x=T/T_{\rm c}$ to an $N$-independent $\hat{x}$. This shift modifies the asymptotic plateau of the aging function from $H(w,x)$ to $H(w,\hat{x})$, in the thermodynamic limit $N \to \infty$. Second, the next-order correction adds a term $\frac{1}{N}  \ln t_{\rm w}$ as $\tilde{x} - \hat{x}\sim \frac{1}{N}  \ln t_{\rm w}$. This correction disappears in the thermodynamic limit, but brings in a $\ln(t_{\rm w})$ decay term to the aging function for a finite $N$, as we show below.

With Eq.~(\ref{eq:Pirelation}),  $\Pi_w^{\rm GTM}(t_{\rm w})$ can be conveniently analyzed using the arcsin law, with  $x$  in Eq.~(\ref{eq:Pi2}) replaced by $\tilde{x}$.
For large $N$, the difference between $\tilde{x}$ and $\hat{x}$,   $\tilde{x} - \hat{x}\sim \frac{1}{N}  \ln t_{\rm w}$, can be considered as a perturbation to the original form. Expanding around $\hat{x}$ to the first order, we obtain, 
\beq\label{eq:Pilogtw}
\begin{split}
    \Pi_w^{\rm GTM}(t_{\rm w})
    &\approx H(w,\tilde{x})\left[1+A\left(\frac{t_{\rm w}}{\tau_0}\right)^{-\alpha} \right]\\
    &\approx H(w,\hat{x})\left[1+A\left(\frac{t_{\rm w}}{\tau_0}\right)^{-\alpha} \right]+\frac{T^2}{Na} \frac{\partial H(w,\hat{x})}{\partial \hat{x}} \ln\left(\frac{t_{\rm w}}{\tau_0}\right)\\
    &=H(w,\hat{x})\left[1+A\left(\frac{t_{\rm w}}{\tau_0}\right)^{-\alpha} \right]- \frac{B}{N}\ln\left(\frac{t_{\rm w}}{\tau_0}\right),
\end{split}
\eeq
where $B= - \frac{T^2}{a} \frac{\partial H(w,\hat{x})}{\partial \hat{x}}$.
The analytic expression of the logarithmic decay $- \frac{B}{N}\ln\left(\frac{t_{\rm w}}{\tau_0}\right)$   with $B= - \frac{T^2}{a} \frac{\partial H(w,\hat{x})}{\partial \hat{x}}$  is compared to the numerical integration of Eq.~(\ref{eq:PI_GTM}) in Fig.~\ref{fig:GTMlogt}: the agreement is converged with increasing $N$. We further compare the theoretical results $- \frac{B}{N}\ln\left(\frac{t_{\rm w}}{\tau_0}\right)$ with the MC data in Fig.~\ref{fig:GREM_nofit}. While the logarithmic form is robust, the coefficient $B$ is not exact due to strong higher-order corrections  for small $N$ in MC simulations. In the main text Fig.~3d, we  treat $B$ as a fitting parameter. 

\begin{figure}
    \centering
    \includegraphics[scale=1]{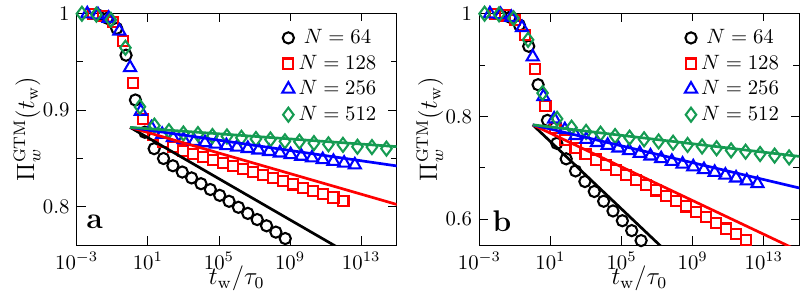}
    \caption{{\bf Logarithmic decay in the GTM at (a) $T=0.64$ and (b) $T=1$.} 
    The data points are obtained by numerical integration of Eq.~(\ref{eq:PI_GTM}). The lines are the $- \frac{B}{N}\ln\left(\frac{t_{\rm w}}{\tau_0}\right)$ term in Eq.~(\ref{eq:Pilogtw}).
    }
    \label{fig:GTMlogt}
\end{figure}

\begin{figure}
    \centering
    \includegraphics[width=0.45\linewidth]{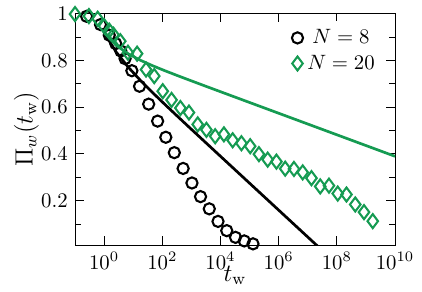}

    \caption{{\bf Comparison between the MC aging function (points) and Eq.~(\ref{eq:Pilogtw}) (lines), at $T=0.75$.}
    No fitting parameters are used for the lines. 
    }
    \label{fig:GREM_nofit}
\end{figure}

\section{Thermodynamics of the finite-size random energy model: a tree-expansion theory}

\subsection{Probability distributions} 

Our aim is to obtain statistical properties of the energy landscape for the finite-sized REM  with $N$ spins, where $N$ is sufficiently small so that the system is far from the thermodynamic limit ($N \to \infty$). We assume that each basin only has one nondegenerate local minimum, and that basins are independent.
The probability distribution of the local minimum $E_{\rm lm}$ is, 
\begin{equation}
\label{eq:pdf-lm}
\begin{split}
p_N^{\rm lm}(E_{\rm lm}) & = (N+1)\rho(E_{\rm lm})\left[ \int_{E_{\rm lm}}^\infty dE \rho(E) \right]^N \\
& = (N+1)\rho(E_{\rm lm})L^N(E_{\rm lm}), 
\end{split}
\end{equation}
where ${L}(E) \equiv\int_E^{\infty} dE \rho(E)$ is  a complementary cumulative distribution function, and $\rho(E)$ is the probability distribution function of the configuration energy $E$. 
The above expression requires that the  $N$ direct neighbors of the local minimum, which are related to the local minimum by a single spin flip,  have an energy $E$ higher  than $E_{\rm lm}$. The factor $N+1$ comes from the $N+1$ choices of the local minimum among the $N+1$ configurations. For the G-REM, $\rho(E) = \rho_{\rm Gauss}(E)$ given by  Eq.~(7), which results in  ${L}(E) = {L}_{\rm Gauss}(E)$, where
\beq
{L}_{\rm Gauss}(E) = \frac12{\rm erfc}(E/\sqrt{2N}),
\eeq
with ${\rm erfc}(x)$ the complementary error function.
For the E-REM,
$\rho(E) = \rho_{\rm exp}(E)$ as in Eq.~(8), and ${L}(E) = L_{\rm exp}(E)$, where
\beq
{L}_{\rm exp}(E) = [1-\exp(E)]\Theta(-E).
\eeq
The probability distributions of the saddle point energy
$E_{\rm sp}$ and the barrier energy $\Delta E$ cannot be explicitly obtained. They need to be computed by 
integrating out the other variable in the  joint distribution $\lambda_N(E_{\rm lm}, E_{\rm sp})$,
\beq
\label{eq:pdf-sp}
    p_N^{\rm sp}(E_{\rm sp})=\int_{-\infty}^{+\infty} dE_{\rm lm}\lambda_N(E_{{\rm lm}},E_{{\rm sp}}),
\eeq
\beq
\label{eq:pdf-be}
    p_N(\Delta E)=\int_{-\infty}^{+\infty} dE_{\rm sp} \lambda_N(E_{\rm sp}-\Delta E,E_{\rm sp}).
\eeq
The joint distribution $\lambda_N(E_{\rm lm}, E_{\rm sp})$ characterizes the  probability of a basin with a local minimum energy  $E_{\rm lm}$ and a saddle point energy $E_{\rm sp}$. Our key task is to compute $\lambda_N(E_{\rm lm}, E_{\rm sp})$. To do that, we develop an approach named {\it tree-expansion theory}. The starting point is to write $\lambda_N(E_{\rm lm}, E_{\rm sp})$ as a summation, 
\beq
\label{eq:febes}
    \lambda_N(E_{\rm lm}, E_{\rm sp}) \equiv \sum_{\Omega=1}^{N} \lambda_N^{(\Omega)}(E_{\rm lm}, E_{\rm sp}),
\eeq
where $\Omega$ is the number of configurations in the basin. For example, $\Omega=1$ means that there is only a local minimum in the basin, and $\Omega=2$ means that there is another configuration in the basin besides the local minimum, etc. The summation converges quickly with the increasing $\Omega$. We next explicitly consider the first three orders, $\Omega = 1, 2, 3$. 

\subsubsection{The first order: $\Omega = 1$}
When $\Omega = 1$, as shown in Fig.~\ref{fig:single_basin_1}, the basin contains only the local minimum without any other configurations. By definition, the saddle point should have the lowest energy among the $N$ neighbors of the local minimum -- the probability of this condition is
${L}^{N-1}(E_{{\rm sp}})$. The saddle point also has $N$ neighbors, including the local minimum and $N-1$ other neighbors. The energies of the other $N-1$  neighbors cannot be all higher than $E_{{\rm sp}}$ -- otherwise the saddle point would be a configuration inside the basin (not a saddle point). This condition imposes a constraint given by $1 - {L}^{N-1}(E_{{\rm sp}})$. Putting these considerations together, we can write the first-order joint distribution as
\beq
    \lambda_{N}^{(1)}(E_{{\rm lm}},E_{{\rm sp}})  =(N+1)N\rho(E_{{\rm lm}})\rho(E_{{\rm sp}})L^{N-1}(E_{{\rm sp}})\left[1-L^{N-1}(E_{{\rm sp}})\right],
\label{eq:1-order}
\eeq
where the extra factor $(N+1)N$ comes from the permutation of choosing a local minimum from the $N+1$ configurations and a saddle point from the rest of the $N$ configurations. 

\begin{figure}
    \centering
    \includegraphics[width=0.3\linewidth]{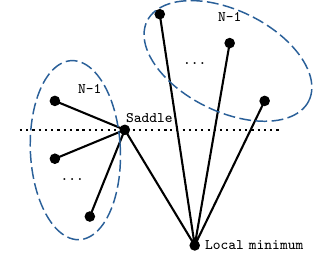}
    \caption{{\bf Graphic representation of the first-order joint distribution Eq.~(\ref{eq:1-order}).}
   Each node represents a configuration, and the height of the node represents the configuration energy $E$. 
   Any pair of configurations connected by a link are related by a single-spin flip.
   The saddle point energy $E_{\rm sp}$ is marked by the dotted line.  For $\Omega =1$, there is only one configuration in the basin ($E<E_{\rm sp}$), which is the local minimum. 
       The number of nodes in the circled cluster is indicated (in this case, both clusters have $N-1$ nodes).      For simplification, only $N=4$ nodes are shown. }
    \label{fig:single_basin_1}
\end{figure}

\subsubsection{The second order: $\Omega = 2$}
\begin{figure}
    \centering
    \includegraphics[width=0.6\linewidth]{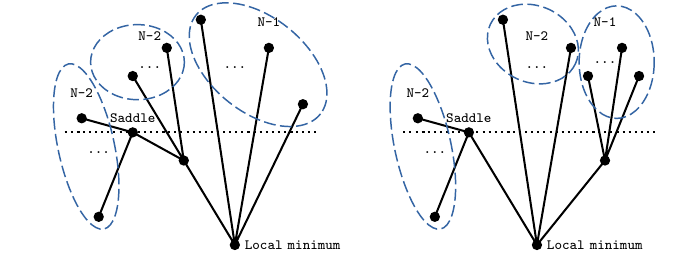}
    \caption{{\bf Graphic representation of the second-order joint distribution Eq.~(\ref{eq:2-order}).}}
    \label{fig:single_basin_2}
\end{figure}

The second-order joint distribution corresponds to two graphs as shown in Fig.~\ref{fig:single_basin_2}. Besides the local minimum, the basin contains one configuration whose energy is between $E_{\rm sp}$ and $E_{\rm lm}$ -- the corresponding probability is $\int_{E_{{\rm lm}}}^{E_{{\rm sp}}}dE\rho(E)$.  The permutation of choosing one saddle point, one local minimum and another configuration in the basin gives a factor of $(N+1)N(N-1)$. The two graphs in Fig.~\ref{fig:single_basin_2} coincidently have the same expression, which gives an extra factor of two. Finally, the second-order joint distribution can be written as, 
\beq
\begin{split}
    \lambda_{N}^{(2)}(E_{{\rm lm}},E_{{\rm sp}}) &  =2(N+1)N(N-1)\rho(E_{{\rm lm}})\rho(E_{{\rm sp}})L^{2N-3}(E_{{\rm sp}})\left[1-L^{N-2}(E_{{\rm sp}})\right]\int_{E_{{\rm lm}}}^{E_{{\rm sp}}}dE\rho(E)\\
     & =2(N+1)N(N-1)\rho(E_{{\rm lm}})\rho(E_{{\rm sp}})L^{2N-3}(E_{{\rm sp}})\left[1-L^{N-2}(E_{{\rm sp}})\right]\left[L(E_{\rm lm})-L(E_{{\rm sp}})\right]
\end{split} 
\label{eq:2-order}
\eeq

\subsubsection{The third order: $\Omega = 3$}
\begin{figure}
    \centering
    \includegraphics[width=0.85\linewidth]{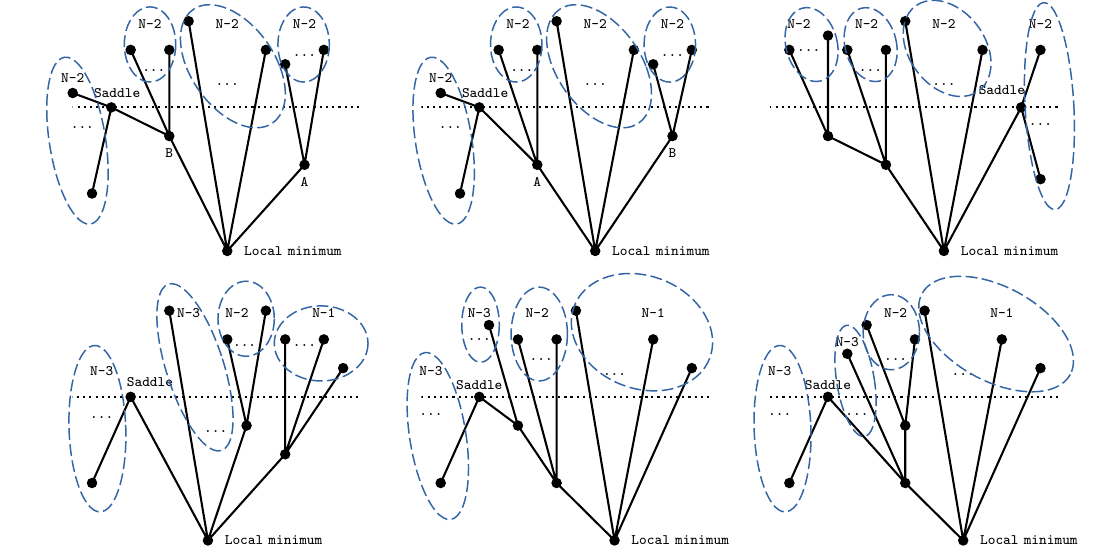}
    \caption{{\bf Graphic representation of the third-order joint distribution Eq.~(\ref{eq:3-order}).}}
    \label{fig:single_basin_3}
\end{figure}

To compute the third-order joint distribution, we need to consider six graphs as shown in Fig.~\ref{fig:single_basin_3}. The top three graphs correspond to the term with $1-L^{N-3}(E_{{\rm sp}})$ in the following expression Eq.~(\ref{eq:3-order}), and the bottom three graphs correspond to the term with $1-L^{N-2}(E_{{\rm sp}})$:
\beq
\begin{split}
 \lambda_{N}^{(3)}(E_{{\rm lm}},E_{{\rm sp}}) & =3(N+1)N(N-1)\rho(E_{{\rm lm}})\rho(E_{{\rm sp}})L^{3N-6}(E_{{\rm sp}}) \times\\
    &\times \left\{(N-2)\left[1-L^{N-3}(E_{{\rm sp}})\right]+(N-1)\left[1-L^{N-2}(E_{{\rm sp}})\right]\right\}\int_{E_{{\rm sp}}>E_{1}>E_{2}>E_{{\rm lm}}}dE_{1}dE_{2}\rho(E_{1})\rho(E_{2})\\
    & =\frac{3}{2}(N+1)N(N-1)\rho(E_{{\rm lm}})\rho(E_{{\rm sp}})L^{3N-6}(E_{{\rm sp}}) \times\\
    &\times \left\{(N-2)\left[1-L^{N-3}(E_{{\rm sp}})\right]+(N-1)\left[1-L^{N-2}(E_{{\rm sp}})\right]\right\}\left[L(E_{\rm lm})-L(E_{{\rm sp}})\right]^{2}.
 \end{split}
 \label{eq:3-order}
 \eeq

\begin{figure}[!htbp]
  \centering
  \includegraphics[scale=0.45]{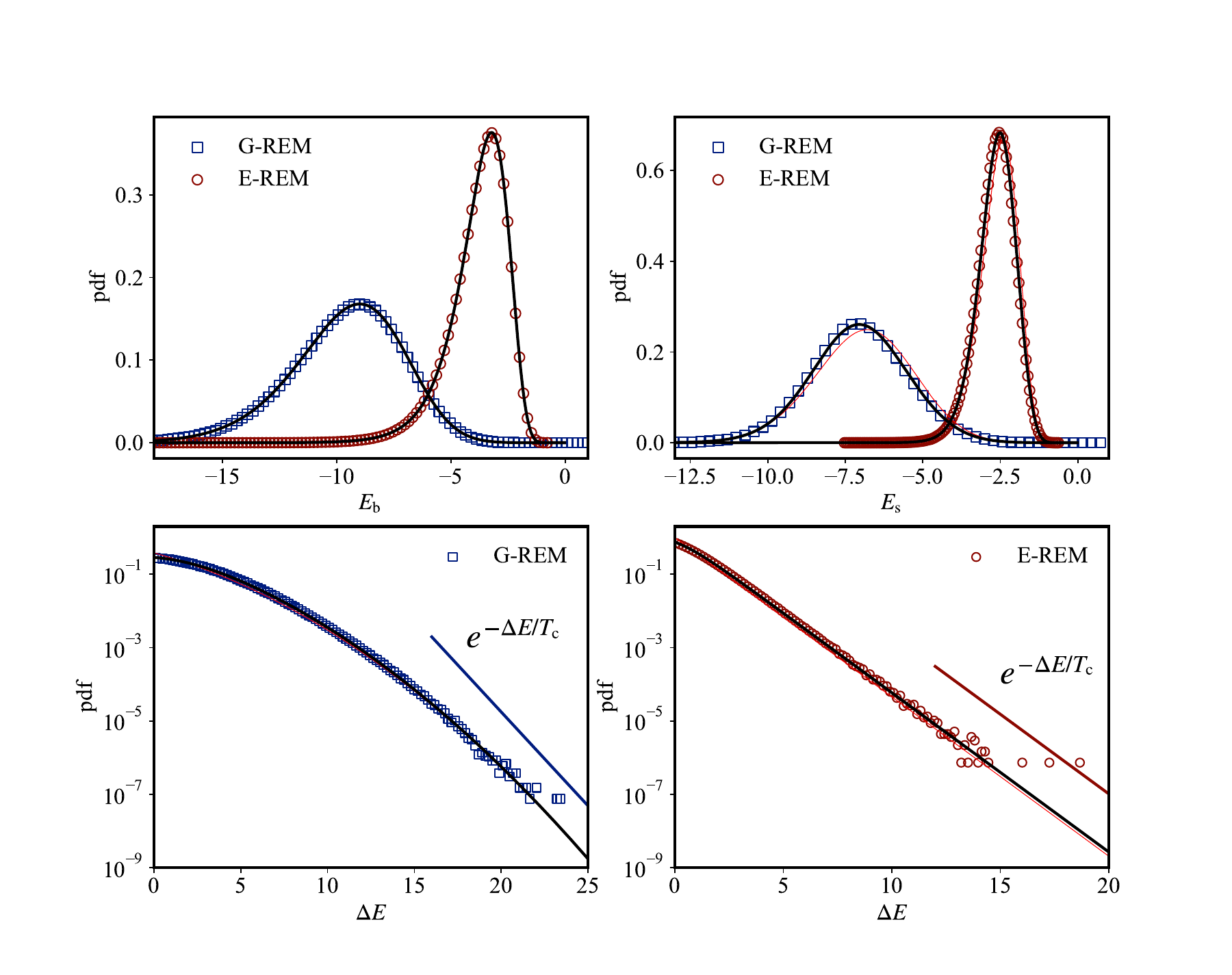}
  \caption{{\bf Comparison between the tree-expansion theory (lines) and the barrier-tree method (points), for the G-REM and E-REM with $N=24$ spins.}
  (a) Probability distribution function (pdf) of the local minimum energy. The line represents Eq.~(\ref{eq:pdf-lm}). (b) Pdf of the saddle point energy. 
  (c,d) Pdf of the barrier energy distribution.  
  In (b-d),  the red line represents  the tree-expansion theoretical results up to the first-order ($\Omega = 1$), and the black line represents the results up to the third-order results ($\Omega = 3$).
  }
  \label{fig:tree_pdfs}
\end{figure}

\subsubsection{Verification with the numerical barrier-tree algorithm}

The above theoretical results are compared with the numerical data obtained by the barrier-tree algorithm for both G-REM and E-REM with $N=24$ spins (Fig.~\ref{fig:tree_pdfs}). 
The expression Eq.~(\ref{eq:pdf-lm}) of the local minimum energy distribution $p_N^{\rm lm}(E_{\rm lm})$ is tested in Fig.~\ref{fig:tree_pdfs}a. The tree-expansion results of the saddle point energy distribution $p_N^{\rm sp}(E_{\rm sp})$ (see Eq.~\ref{eq:pdf-sp}) and the barrier energy distribution  $p_N(\Delta E)$ (see Eq.~\ref{eq:pdf-be}) are tested in Fig.~\ref{fig:tree_pdfs}b-d. It can be seen that the first-order results are already very close to the numerical data. Note that the distribution $\rho(E)$ in the theoretical expressions shall be replaced by Eqs.~(7) and (8) in the Appendix for the corresponding REMs.

\begin{figure}[!htbp]
  \centering
  \includegraphics[scale=0.8]{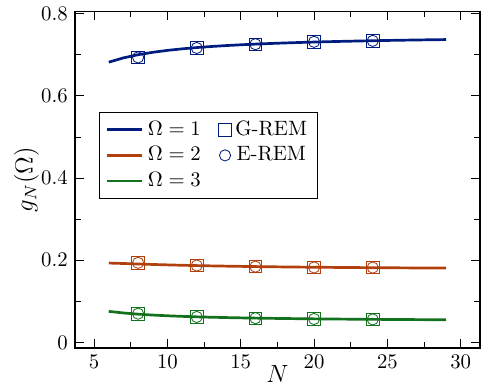}
  \caption{{\bf Probability $g_N(\Omega)$ of having $\Omega$ configurations in a basin}.
  Lines are tree-expansion theory results Eq.~(\ref{eq:pb1}), and points are numerical data obtained by the barrier-tree algorithm.
  }
  \label{fig:pBasin-N}
\end{figure}

\subsection{Internal entropy}\label{sec:internal_entropy}

The probability  of having $\Omega$ configurations in a basin is, 
\beq\label{eq:pb}
g_N(\Omega)=\int_{-\infty}^{\infty} dE_{\rm sp} \int_{-\infty}^{E_{\rm sp}} dE_{\rm lm} \lambda_N^{(\Omega)}(E_{\rm lm},E_{\rm sp}).
\eeq
Using the above tree-expansion results, Eqs.~(\ref{eq:1-order}),~(\ref{eq:2-order}) and (\ref{eq:3-order}), we obtain the first three orders, 
\beq\label{eq:pb1}
\begin{split}
g_N(1)
    &= \int_{-\infty}^{\infty} dE_{\rm sp} \rho(E_{\rm sp}) N(N+1) L^{N-1}(E_{\rm sp})\left[1-L^{N-1}(E_{\rm sp})\right] \left[1-L(E_{\rm sp})\right] \\
    &= \int_0^1 d\sigma N(N+1)\sigma^{N-1}(1-\sigma^{N-1})(1-\sigma)\\
    &= \frac{3N-3}{4N-2}\\  
    &\overset{N \to \infty }{=}  \frac{3}{4},\\
g_N(2) &= \int_0^1 d\sigma (N-1)N(N+1)\sigma^{2N-3}(1-\sigma^{N-2})(1-\sigma)^2\\
&= \frac{(N+1)(N-2)(19N-12)}{6(2N-1)(3N-4)(3N-2)}\\
&\overset{N \to \infty }{=}  \frac{19}{108},\\
g_N(3) &= \int_0^1 d\sigma (N-1)N(N+1)\sigma^{3N-6}
\left[\frac{N-2}{2}(1-\sigma^{N-3})+\frac{N-1}{2}(1-\sigma^{N-2})\right]
(1-\sigma)^3,\\
&={\frac{N(N+1)(2N-3)}{(3N-5)(3N-4)(3N-2)}-\frac{3(N-1)N(N+1)}{4(2N-3)(4N-7)(4N-5)}}\\
&\overset{N \to \infty }{=}  \frac{175}{3456}.
\end{split}
\eeq
Note that the results in Eq.~(\ref{eq:pb1}) are model-independent, i.e., independent of $\rho(E)$. The numerical data obtained by the barrier-tree algorithm confirm the tree-expansion theory Eq.~(\ref{eq:pb1}) (see Fig.~\ref{fig:pBasin-N}).

In the thermodynamic limit ($N \to \infty$), the probability $g_N(\Omega)$ converges to an analytical expression that is universal for any $\Omega$,
\beq
\begin{split}\label{eq:P-Omega}
g_\infty(\Omega) &= \lim_{N\rightarrow\infty}\int_0^1 d\sigma \frac{N!}{(N-\Omega-1)!}\sigma^{\Omega N-\frac12 \Omega(\Omega+1)}(1-\sigma^{N-\Omega})(1-\sigma)^\Omega\\
&=\lim_{N\rightarrow\infty}\frac{N!\Omega!}{(N-\Omega-1){!}}
\left[
\frac{(\Omega N-\frac12 \Omega(\Omega+1))!}{(\Omega N-\frac12 \Omega(\Omega+1)+1)!} 
- \frac{((\Omega+1)N-\frac12 \Omega(\Omega+3))!}{((\Omega+1)N-\frac12 \Omega(\Omega+1)+1)!} 
\right] \\
&=\frac{\Omega!}{\Omega^{\Omega+1}}-\frac{\Omega!}{(\Omega+1)^{\Omega+1}}.
\end{split}
\label{eq:p_omega}
\eeq

\begin{figure}[!htbp]
  \centering
  \includegraphics[width=0.6\linewidth]{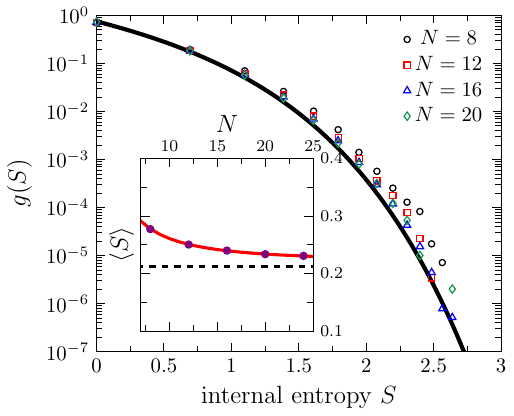}
  \caption{{\bf Probability distribution function of the internal entropy for the G-REM.} 
   The data points are $g^{\rm BT}(S)$ obtained by the numerical barrier-tree method, and  
the solid line is Eq.~(\ref{eq:p_omega}).
  (inset) Average numerical entropy $\langle S \rangle$ as a function of $N$, where the dashed line is $S_\infty \approx 0.212439 $ (Eq.~\ref{eq:internal_entropy}).
  }
  \label{fig:entropy}
\end{figure}

The internal entropy of the whole system in the thermodynamic limit is
\beq\label{eq:internal_entropy}
S_\infty \equiv \sum_{\Omega=1}^\infty g_\infty(\Omega)\log\Omega \approx 0.212439.
\eeq
This result is consistent with the numerical data obtained by the barrier-tree method (see Fig.~~\ref{fig:entropy}). 
The inset of Fig.~\ref{fig:entropy} shows that the mean internal entropy $\langle S \rangle$  converges to a value $S_\infty  \approx 0.21$ in the large-$N$ limit. Then the average number of configurations in a basin is $\Omega \approx e^{0.21} \approx 1.2 $, i.e., each basin contains approximately one configuration in average. 
Thus in the spin glass phase, the mean internal entropy per spin vanishes ($\langle S \rangle/N \to 0$) in the thermodynamic limit, consistent with thermodynamic  theory of the REM~\cite{derrida1980random}.

\subsection{Number of basins}\label{sec:Nb}
We assume that each basin contains one local minimum. Two local minima cannot be directly related via a single-spin flip -- otherwise the two basins cannot be separated. Thus, in average, we can assume that each basin occupies $N+1$ configurations, i.e., one local minimum and its $N$ neighbours. The total number of configurations for  a $N$-spin REM is $2^N$.  
Therefore, the total number of basins is, 
\beq\label{eq:Nb-N}
N_{\rm b} = \frac{2^N}{N+1}.
\eeq
Figure~\ref{fig:Nb} shows that this simple consideration Eq.~(\ref{eq:Nb-N}) describes well the numerical data obtained by the barrier-tree method.

\begin{figure}[!htbp]
  \centering
  \includegraphics[scale=1]{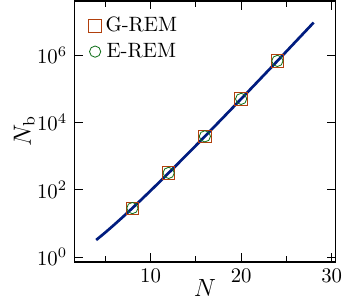}
  \caption{{Number of basins $N_{\rm b}$ as a function of the system size $N$}.
  The line represents Eq.~(\ref{eq:Nb-N}), and the points are data obtained by the numerical barrier-tree algorithm. 
  }
  \label{fig:Nb}
\end{figure}


\end{document}